\documentclass[twocolumn]{aastex631}
\usepackage{lineno}

\newcommand\aastex{AAS\TeX}

\usepackage{xcolor}
\usepackage{wrapfig}
\usepackage{url}
\usepackage{multirow}
\usepackage{booktabs} 
\usepackage{threeparttable}
\usepackage{natbib}

\usepackage[T1]{fontenc} 
\usepackage[utf8]{inputenc}  
\usepackage[graphicx]{realboxes} 

\usepackage{tipa}

\shorttitle{\aastex\ "Mega-Earths" are Likely an Artifact of Stellar Activity}
\shortauthors{Brinkman et al.}

\begin{document}
\newcommand{\PSUAA}{Department of Astronomy \& Astrophysics, 525 Davey Laboratory, The Pennsylvania State University, University Park, PA, 16802, USA}
\newcommand{\PSUCEHW}{Center for Exoplanets \& Habitable Worlds, 525 Davey Laboratory, The Pennsylvania State University, University Park, PA 16802, USA}
\newcommand{\PSETI}{Penn State Extraterrestrial Intelligence Center, 525 Davey Laboratory, The Pennsylvania State University, University Park, PA 16802, USA}
\newcommand{\UA}{Steward Observatory, The University of Arizona, 933 N.\ Cherry Ave, Tucson, AZ 85721, USA}
\newcommand{\Caltech}{Department of Astronomy, California Institute of Technology, Pasadena, CA 91125, USA}
\newcommand{\JHU}{Department of Physics \& Astronomy, Bloomberg Center, Johns Hopkins University, Baltimore, MD 21218, USA}
\newcommand{\Macquarie}{School of Mathematical and Physical Sciences, Macquarie University, Balaclava Road, North Ryde, NSW 2109, Australia}
\newcommand{\CUBoulder}{Department of Physics, 390 UCB, University of Colorado, Boulder, CO 80309, USA}
\newcommand{\JPL}{Jet Propulsion Laboratory, California Institute of Technology, 4800 Oak Grove Drive, Pasadena, CA 91109, USA}
\newcommand{\MITEAPS}{Department of Earth, Atmospheric, and Planetary Sciences, Massachusetts Institute of Technology, Cambridge, MA 02139, USA}
\newcommand{\MITKavli}{Kavli Institute for Astrophysics and Space Research, Massachusetts Institute of Technology, Cambridge, MA 02139, USA}
\newcommand{\UCI}{Department of Physics \& Astronomy, The University of California, Irvine, Irvine, CA 92697, USA}
\newcommand{\Carnegie}{Earth and Planets Laboratory, Carnegie Institution for Science, 5241 Broad Branch Road, NW, Washington, DC 20015, USA}
\newcommand{\PSUICS}{Institute for Computational and Data Sciences, The Pennsylvania State University, University Park, PA 16802, USA}
\newcommand{\PSUCASt}{Center for Astrostatistics, 525 Davey Laboratory, The Pennsylvania State University, University Park, PA 16802, USA}
\newcommand{\Princeton}{Department of Astrophysical Sciences, Princeton University, 4 Ivy Lane, Princeton, NJ 08540, USA}
\newcommand{\IAS}{Institute for Advance Study, 1 Einstein Drive, Princeton, NJ 08540, USA}
\newcommand{\Tsinghua}{Department of Astronomy, Tsinghua University, Beijing 100084, China}
\newcommand{\FlatironCCA}{Center for Computational Astrophysics, Flatiron Institute, 162 Fifth Avenue, New York, NY 10010, USA}
\newcommand{\ETH}{ETH Zurich, Institute for Particle Physics \& Astrophysics, Zurich, Switzerland}
\newcommand{\UCO}{UC Observatories, University of California, Santa Cruz, CA 95064, USA}
\newcommand{\SantaCruz}{University of California, Santa Cruz}
\newcommand{\WMKO}{W.\ M.\ Keck Observatory, 65-1120 Mamalahoa Hwy, Kamuela, HI 96743, USA}
\newcommand{\SSL}{Space Sciences Laboratory, University of California, Berkeley, CA 94720, USA}
\newcommand{\UH}{Institute for Astronomy, University of Hawai‘i, 2680 Woodlawn Drive, Honolulu, HI 96822, USA}
\newcommand{\UCB}{Department of Astronomy, 501 Campbell Hall, University of California, Berkeley, CA 94720, USA}
\newcommand{\UCLA}{Department of Physics \& Astronomy, University of California Los Angeles, Los Angeles, CA 90095, USA}
\newcommand{\nexsci}{NASA Exoplanet Science Institute/Caltech-IPAC, MC 100-22, 1200 E.\ California Blvd., Pasadena, CA
91125, USA}
\newcommand{\COO}{Caltech Optical Observatories, California Institute of Technology, Pasadena, CA 91125, USA}
\newcommand{\Sydney}{Sydney Institute for Astronomy (SIfA), School of Physics, University of Sydney, NSW 2006, Australia}
\newcommand{\Kansas}{Department of Physics and Astronomy, University of Kansas, Lawrence, KS, USA}
\newcommand{\Warwick}{Physics Department, University of Warwick, Coventry CV4 7AL, United Kingdom}
\newcommand{\Yale}{Department of Astronomy, Yale University, New Haven, CT 06520, USA}
\newcommand{\ICL}{Astrophysics Group, Department of Physics, Imperial College London, Prince Consort Rd, London SW7 2AZ, UK}
\newcommand{\Schmidt}{Astrophysics \& Space Institute, Schmidt Sciences, New York, NY 10011, USA}
\newcommand{\Amsterdam}{University of Amsterdam}
\newcommand{\ND}{Department of Physics and Astronomy, University of Notre Dame, Notre Dame, IN 46556, USA}
\newcommand{\Geneva}{Observatoire Astronomique de l'Université de Genève, Chemin Pegasi 51, 1290 Versoix, Switzerland}

\title{"Mega-Earths" are likely an artifact of stellar activity}

\correspondingauthor{Casey L. Brinkman}
\email{casey.brinkman@mcgill.ca}

\author[0000-0002-4480-310X]{Casey L. Brinkman}
\altaffiliation{TSI Postdoctoral Fellow}
\affiliation{McGill University, Trottier Space Institute, 3550 rue University, Montreal, QC, H3A 2A7, CAN}

\author[0000-0001-6129-5699]{Nicolas B. Cowan}
\affiliation{McGill University, Trottier Space Institute, 3550 rue University, Montreal, QC, H3A 2A7, CAN}

%\author[]{Nic Cowan}

%\author[]{} Andrew Cumming? ChEd? Sabina?

\begin{abstract}

Empirical relations and planet formation simulations suggest an upper mass limit for rocky planets of $\sim$10M$_{\oplus}$, as cores above this will almost certainly undergo runaway gas accretion. Nevertheless, a small number of planets appear to have densities consistent with rocky Earth-like compositions above this threshold, indicating that unusual formation and evolutionary scenarios are capable of producing extremely massive rocky planets called "Mega-Earths." However, all of these planets orbit active stars with log$R'_{HK}$ $>$-4.75. We find a <1$\%$ probability that all literature ultra-massive rocky planets would coincidentally orbit such active stars, and find no astrophysical reason to expect Mega-Earths to preferentially form around active stars. We therefore conclude that Mega-Earths are artifacts of stellar activity, and we question the reliability of Gaussian Processes to enable mass measurements for rocky planets around active stars.

\end{abstract}


\section{Introduction}
\label{sec:intro}
The diversity of rocky planet masses, radii, and densities inform our understanding of planet formation and evolution pathways. Rocky planets can either be formed primordially \citep{2014ApJ...797...95L, 2020AJ....159..211C}, or as the stripped cores of larger sub-Neptunes whose envelopes were likely lost through core-powered mass loss \citep{2018MNRAS.476..759G}, photoevaporation from their host star \citep{2017ApJ...847...29O}, or a combination of phenomena \citep{2021MNRAS.508.5886R}. The transition from rocky to gas-rich planets is a natural product of core-accretion theory: once a growing planetary core exceeds several Earth masses, it can no longer avoid runaway gas accretion from the surrounding protoplanetary disk, whereby massive planets necessarily form gaseous envelopes \citep{1980PThPh..64..544M}. 

Empirically, the mass and radius measurements of small planets suggest a transition  between planets consistent with a bare rocky composition devoid of significant volatile content, and those likely hosting envelopes at $R=1.5-1.75 R_{\oplus}$ \citep{2014ApJ...783L...6W, 2014ApJ...792....1L, 2015ApJ...801...41R, 2017AJ....154..109F}. This transition point in planet mass occurs between $M=4-5 M_{\oplus}$, and appears to be the minimum mass needed to accrete substantial quantities of volatile elements \citep{2020A&A...634A..43O, 2024A&A...686A.296M}. 

This empirical transition happens at masses far lower than the theoretical limit derived from planet formation theories. Simulations of collisional mantle stripping and post-formation erosion suggest that even under extreme conditions, the maximum mass for a primordial rocky planet is unlikely to exceed $\sim$10 M$_{\oplus}$ \citep{1996Icar..124...62P, 2025ApJ...980L..40L, 2022MNRAS.517.3132R}. This cutoff boundary is a function of disk composition, stellar flux, etc. The existence of an ultra-massive rocky planet beyond this threshold would require either exceptional formation environments or extreme evolutionary pathways, and would imply interior conditions potentially unlike other rocky worlds. 

In recent years, several exoplanet characterization efforts have yielded mass and radius measurements consistent with volatile-poor densities, but with masses larger than 10 M$_{\oplus}$, occasionally referred to as "Mega-Earths." These planets could potentially form collisionally post disk-dispersal thereby escaping runaway gas accretion \citep{2015ApJ...811...41L}, or they could be remnant photo-evaporated cores of former giants \citep{2020Natur.583...39A}. Our certainty that these strange objects even exist, however, is only as strong as the quality of the mass and radius measurements and therefore our ability to characterize the host star.
\begin{figure*}
    \centering
    \includegraphics[width=0.9\linewidth]{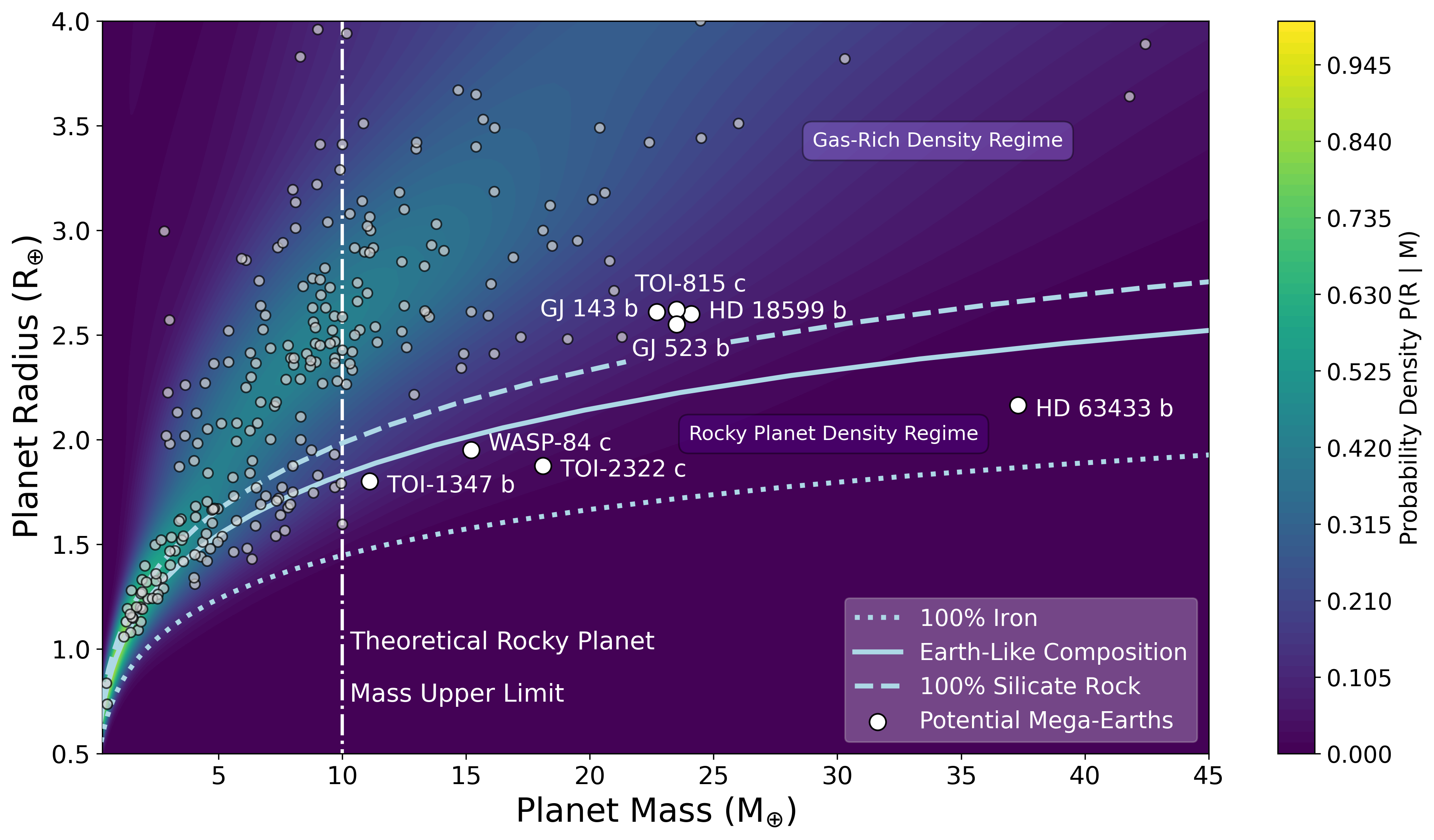}

    \caption{Planet radius vs.\ mass for small planets ($ < 4R_{\oplus}$) are plotted as grey points. Lines show mass-radius relations for planets with an Earth-like composition, a purely silicate-rock composition, and a purely iron composition \citep{2019PNAS..116.9723Z}, representing the possible mass-radius space in which rocky planets can exist. Literature values come from the NASA Exoplanet Archive  \citep{2013PASP..125..989A} queried 7/15/2026 for planets with mass and radius measurements with fractional uncertainties $<$30$\%$. The colormap represents the normalized probability density estimated from a Gaussian KDE. We show the theoretical maximum rocky planet mass threshold (10 M$_{\oplus}$) and identify planets above this threshold in the rocky planet density regime as possible Mega-Earths(white circles). }
    \label{fig:massrad}
\end{figure*}  

Magnetically active stars notoriously complicate stellar characterization and challenge RV mass determination. Active regions can suppress convection in the star, producing erroneous RV measurements and are capable of masking the smaller semi-amplitudes produced by small, rocky planets. Star spots (and plages) can mimic the signature of an orbiting planet as the star rotates, and this added jitter can artificially inflate the RV signal of a transiting planet or bury it beneath noise, often leading to erroneously large planet masses \citep{dumusque2011two, dumusque2011one}. 

In this work, we use literature masses and radii of small planets to identify potential "Mega-Earth" planets beyond the runaway gas accretion threshold. We then analyze the likelihood of their existence given the context of their host star activity. 
%Thankfully, we can employ two tools not used in the prior RV analysis to measure a more accurate and precise planet mass: high-cadence observations and Gaussian Process fitting. The likely rotational period of WASP-84 ($\sim$14 days) is an order of magnitude longer than the orbital period of WASP-84 c (1.45 days), therefore with high cadence RV sampling over a small number of observing nights, we can fully sample the phase of the planet with minimal impact from stellar activity. Additionally, with the TESS lightcurve we can acquire strong physically motivated priors on kernel parameters, allowing us to account for stellar variability using Gaussian Processes \citep{2014MNRAS.443.2517H, 2023AJ....165...74B}. \textbf{With 4 hours on Gemini-N/MAROON-X we will reduce the 1$\sigma$ mass uncertainty from 4.5 M$_{\oplus}$ to 0.6 M$_{\oplus}$, allowing us to determine if WASP-84 c is the first bona-fide Ultra-Massive Rocky Planet.} This mass precision will yield a constraint on the interior iron fraction of 10$\%$, enabling us to investigate the potentially non-Earthlike equations of state present in the interior \citep{2013PhRvB..87i4102B, PhysRevResearch.5.033194}. While the discovery of an ultra-massive rocky planet will require creative solutions to avoid runaway gas accretion, the absence of such planets tells us that these pathways--such as collisions post disk-dispersal or atmosphere stripping--are widely unable to produce enlarged rocky planets. 

\section{Mega-Earths and the Small Planet Population}

\begin{figure*}
    \centering
    \includegraphics[width=0.4\linewidth]{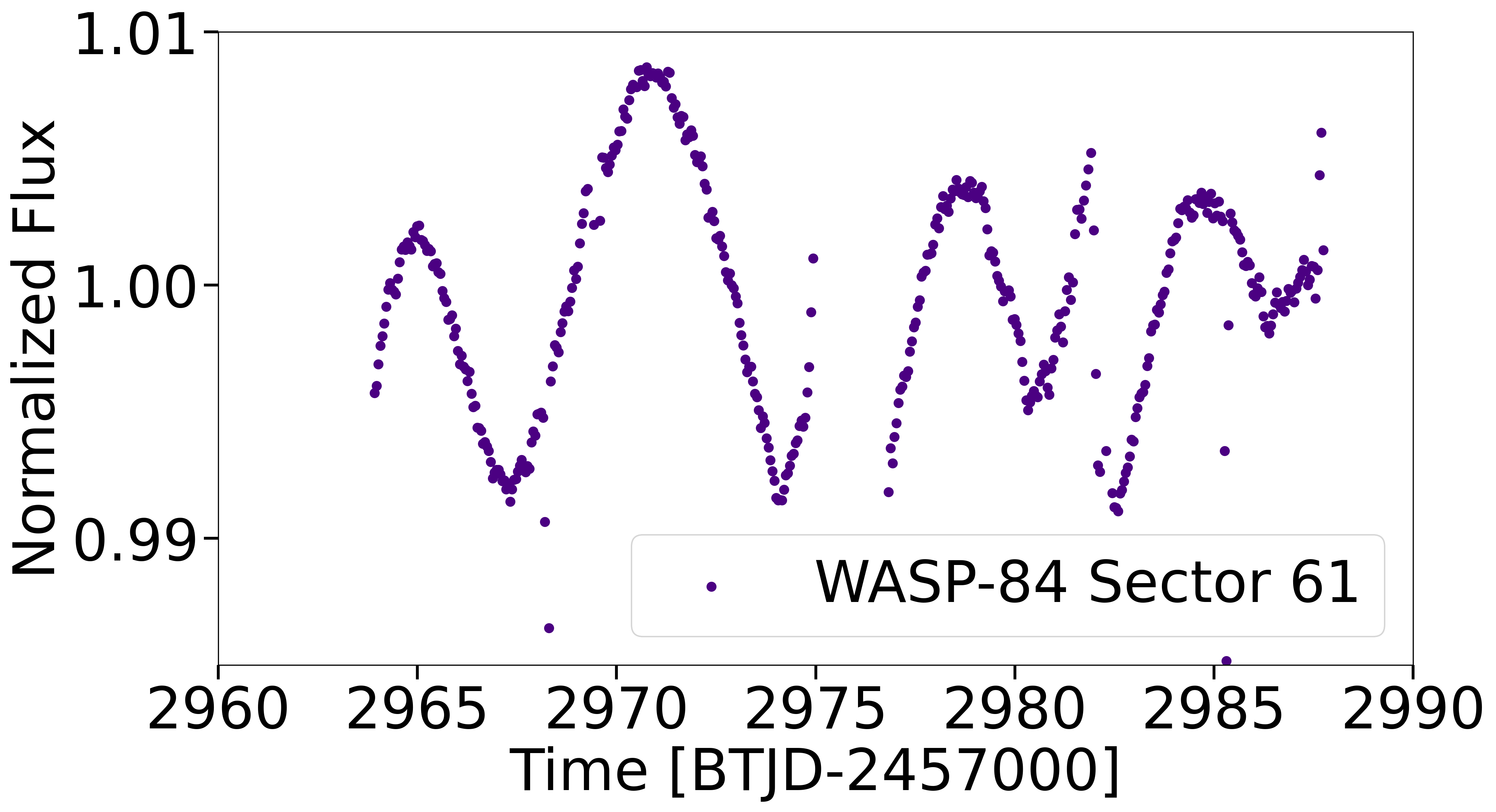}
    \includegraphics[width=0.4\linewidth]{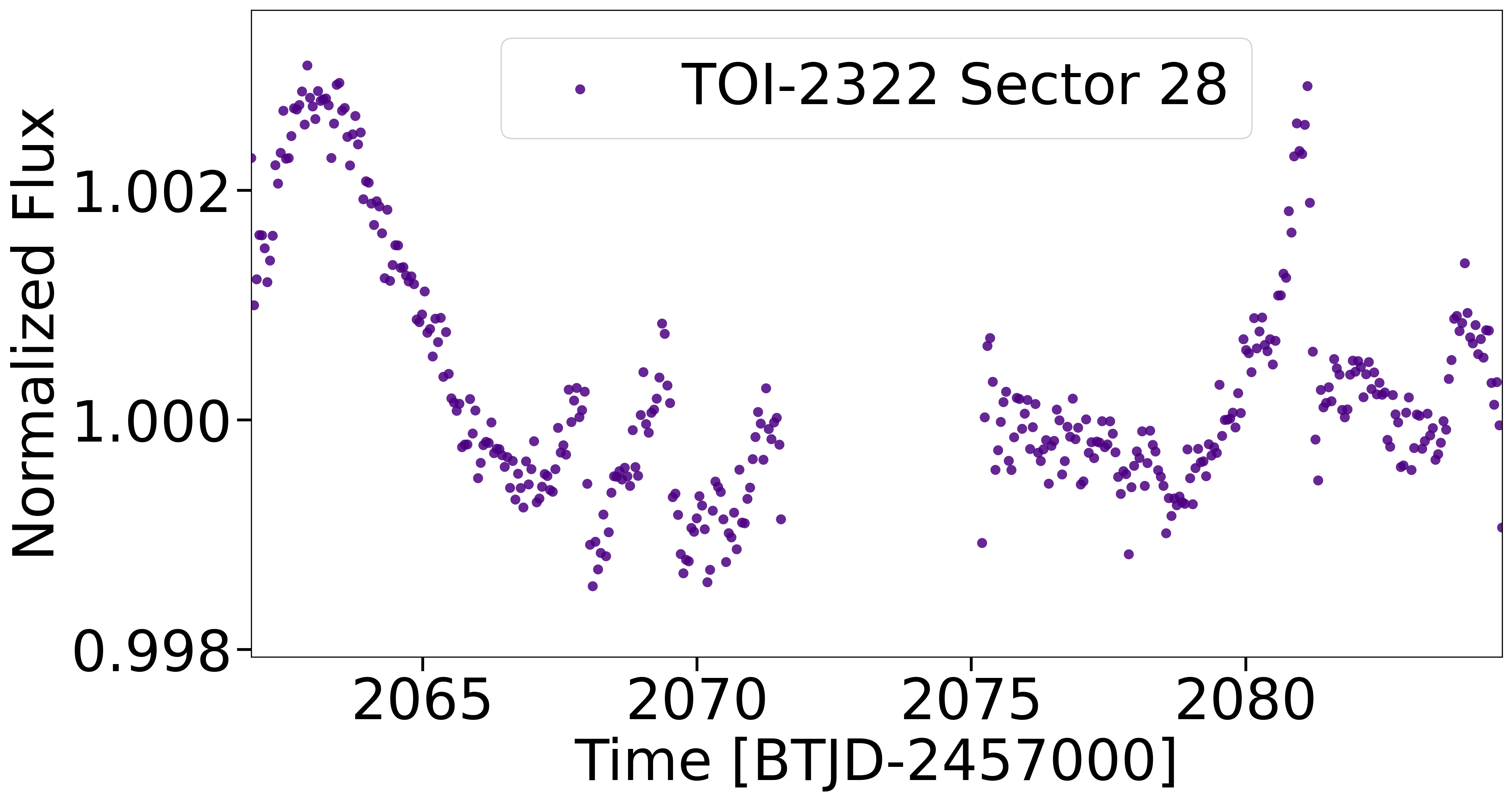}
    \includegraphics[width=0.4\linewidth]{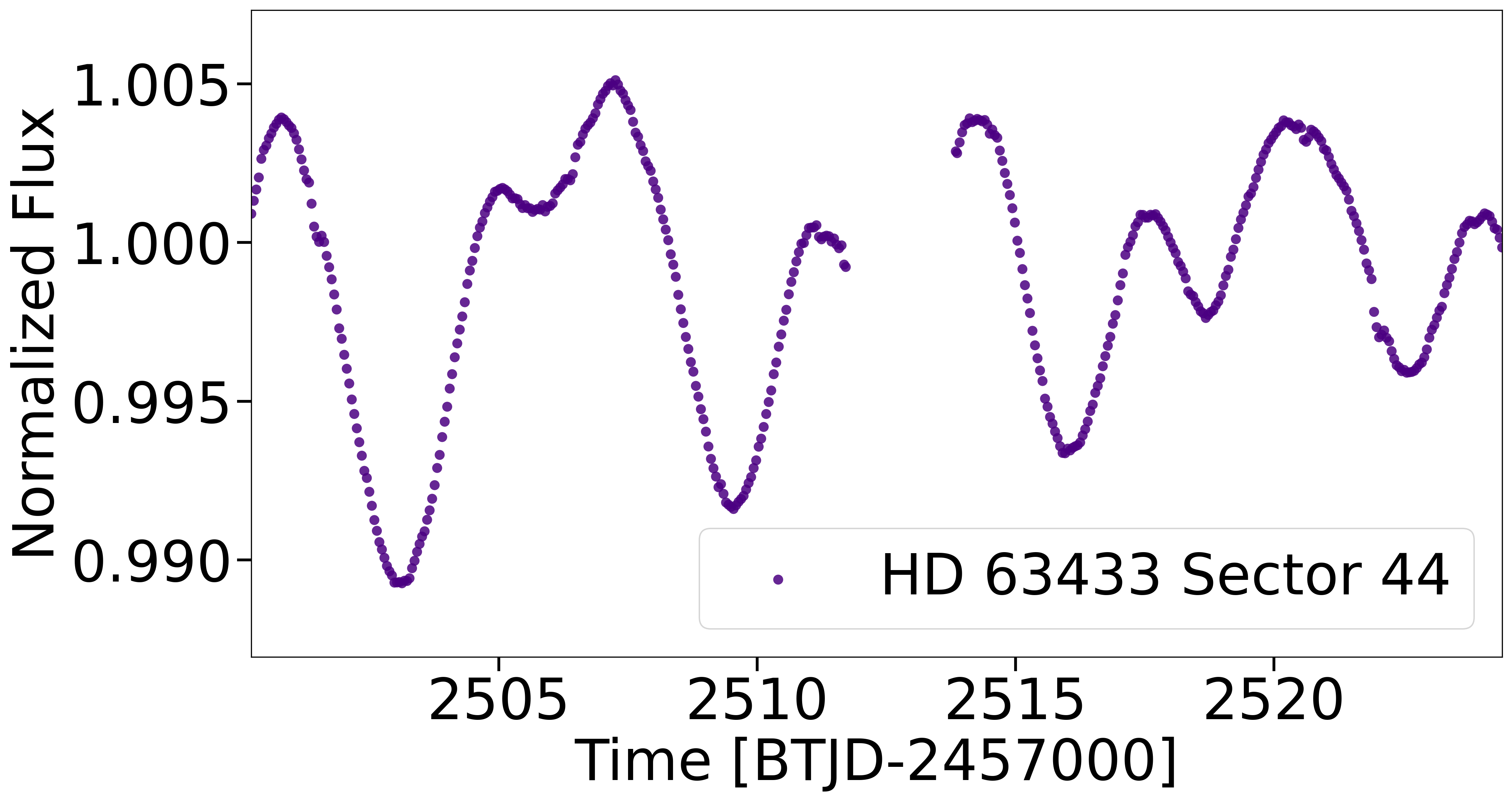}
    \includegraphics[width=0.4\linewidth]{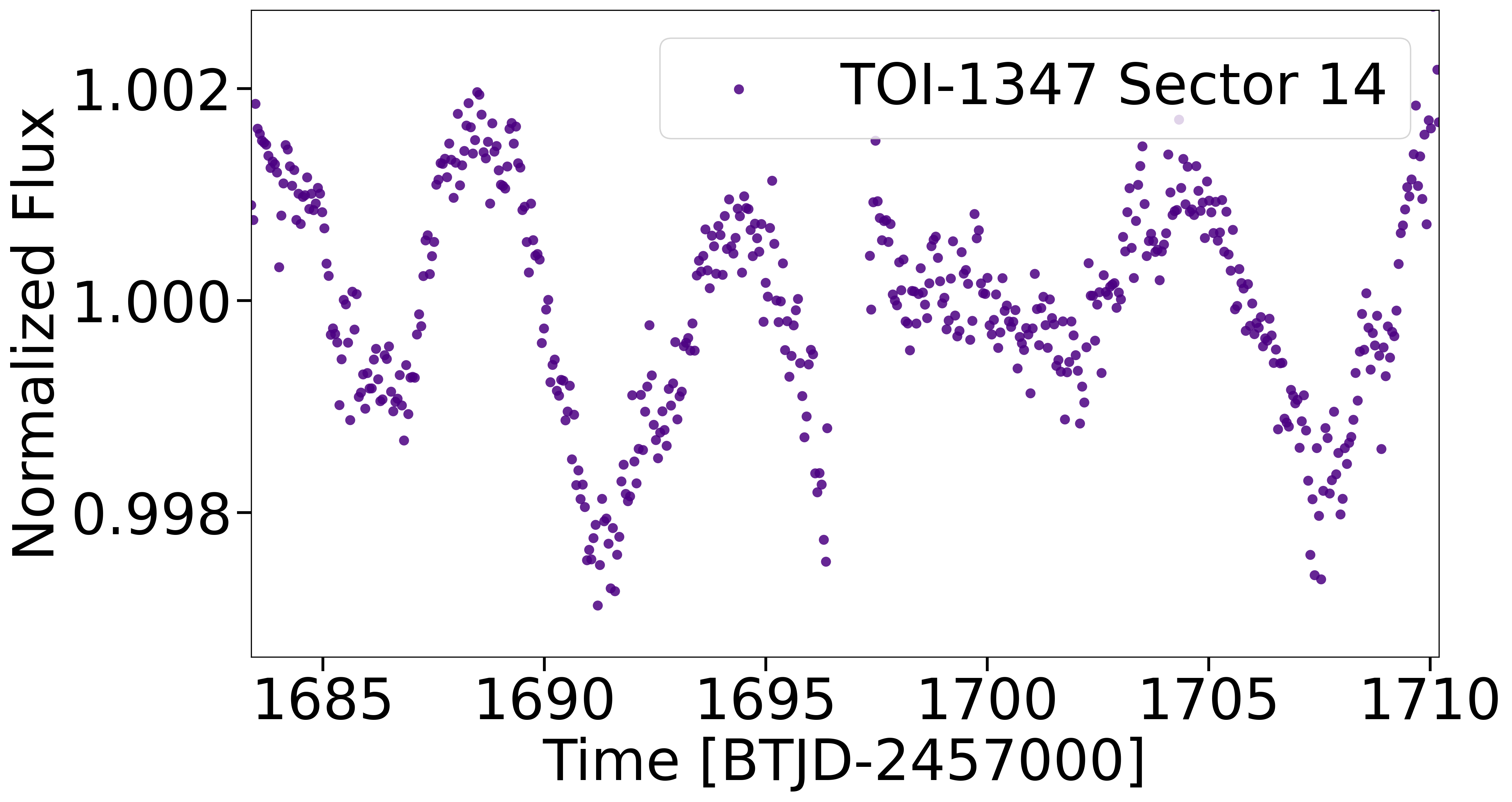}
    \includegraphics[width=0.4\linewidth]{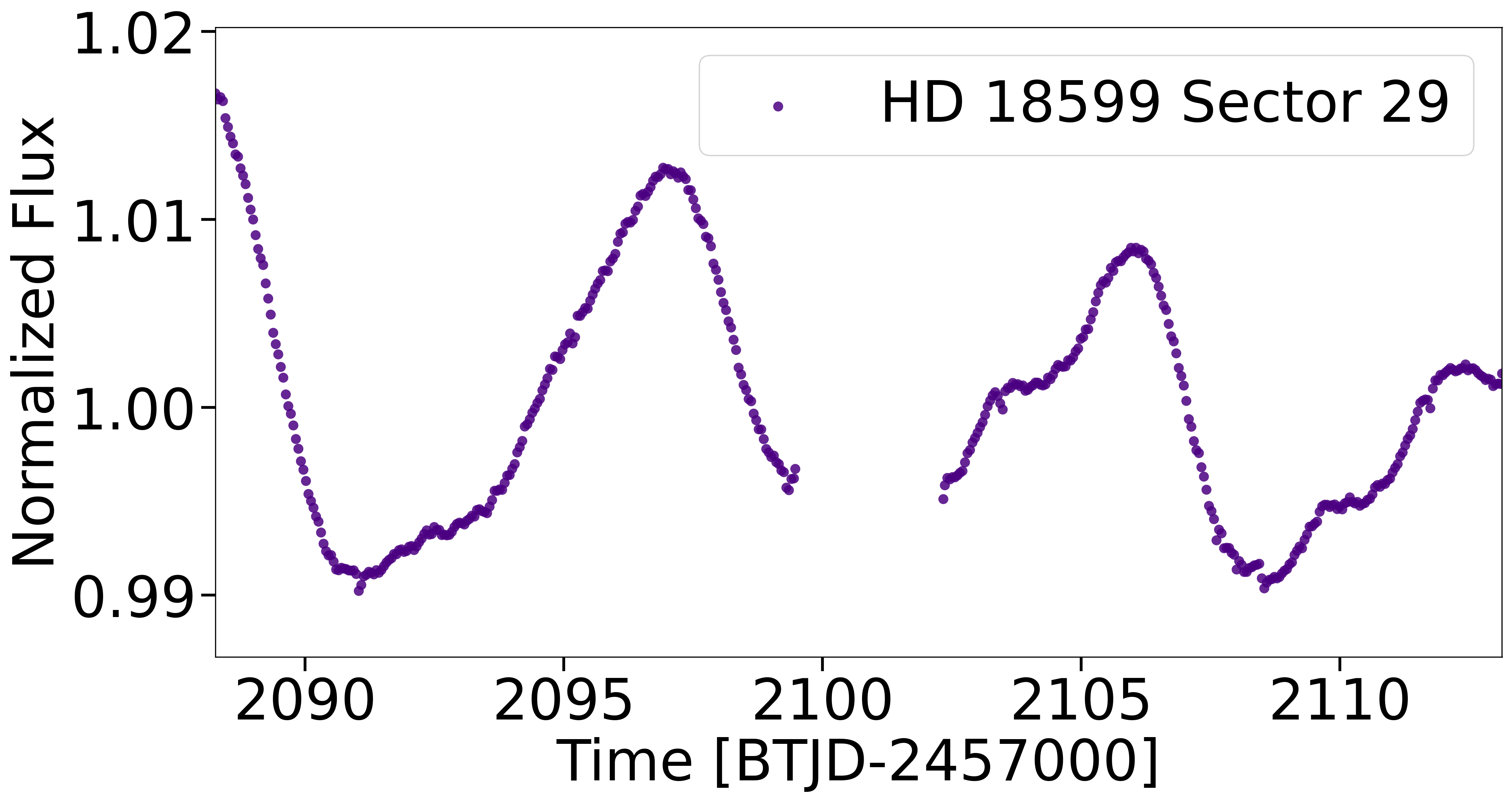}
    \includegraphics[width=0.4\linewidth]{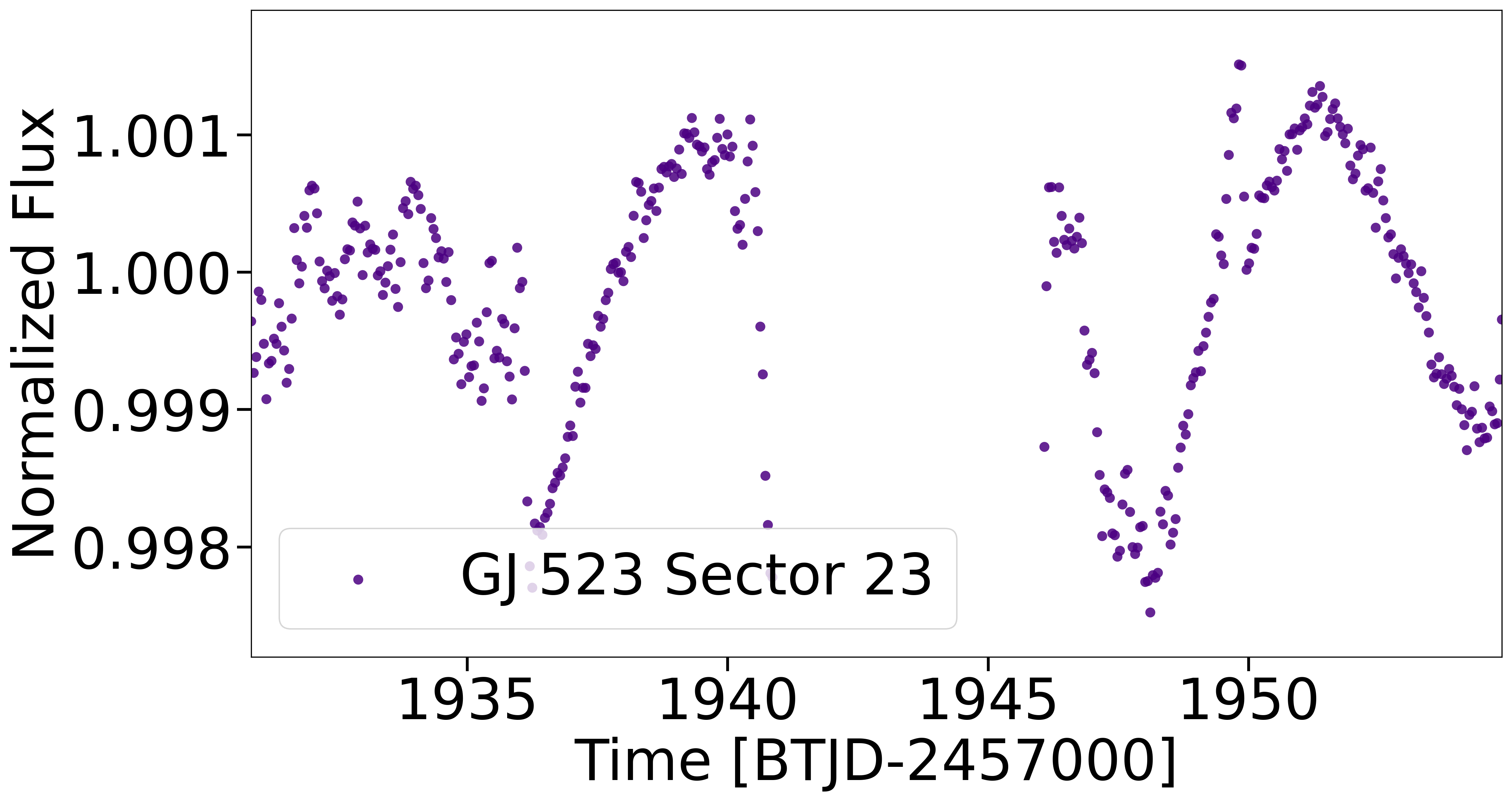}
    \includegraphics[width=0.4\linewidth]{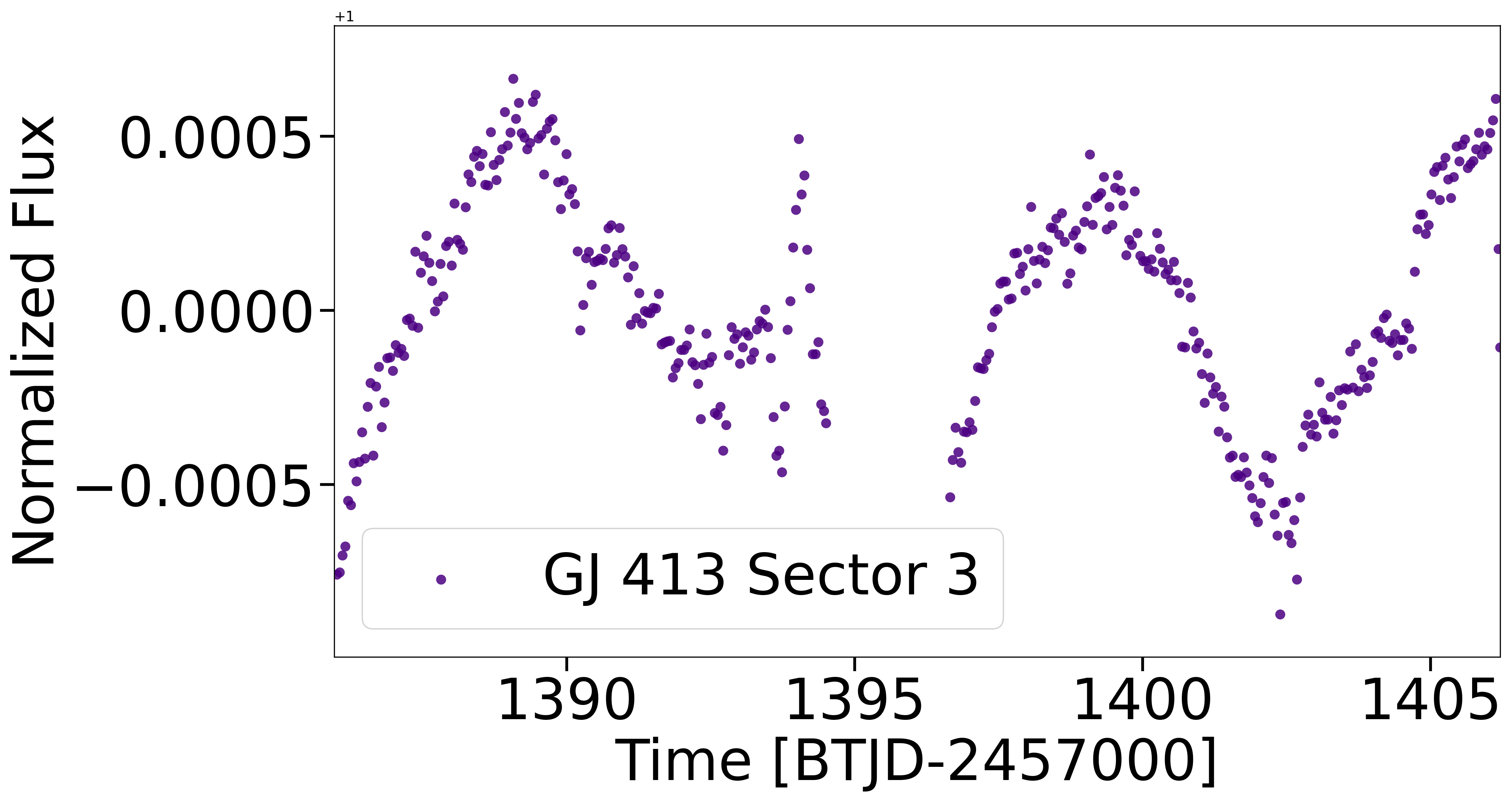}
    \includegraphics[width=0.4\linewidth]{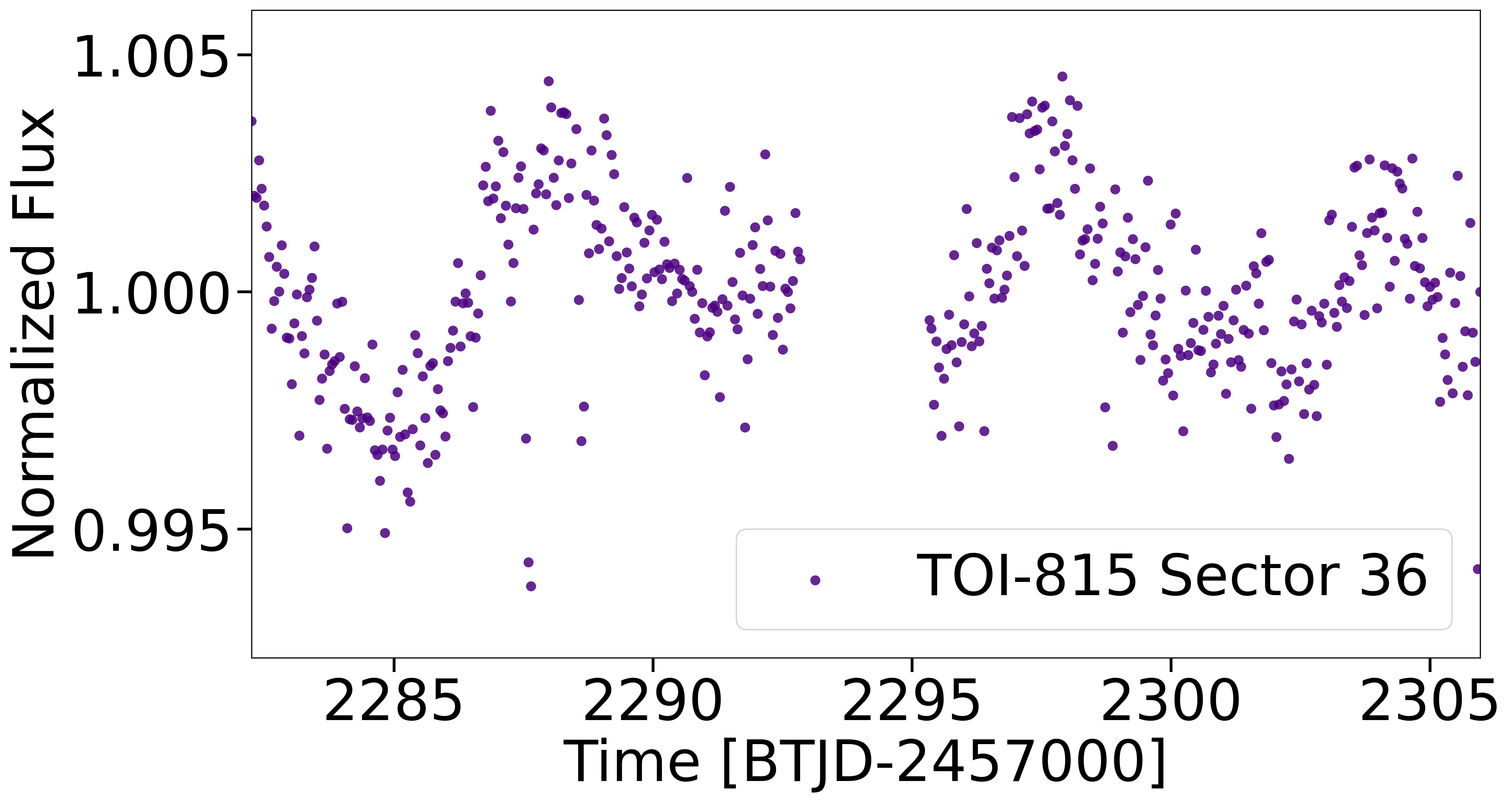}

    \caption{TESS lightcurves (PDCSAP Flux) of each Mega-Earth host star are shown, demonstrating visible rotational modulation due to stellar activity.}
    \label{fig:lightcurves}
\end{figure*}
\subsection{Sample Curation}
To identify planets in the rocky planet density regime, we selected planets between 0--4 R$_{\oplus}$ that have published masses and radii using the NASA Exoplanet Archive \citep[queried 7/15/2025]{2013PASP..125..989A}. First, we removed any planets with only mass upper limits, nontransiting m$\sin(i)$ values, or masses produced using mass-radius relationships. Next, we removed planets with fractional uncertainties on mass larger than 50$\%$. Lastly, we removed planets with masses measured using transit timing tariations (TTVs), as TTVs---particularly from large surveys---often predict unphysically high planet masses that are unable to be validated through other methods, and are more reliable for mass ratios between planets than absolute masses \citep{2017ApJ...839L...8M, 2026ApJ..1005...37J}. The radii and radial velocity (RV) masses of our resulting sample of 333 planets are shown in Figure \ref{fig:massrad}. 

\subsection{Anomalously Massive Rocky Planets}

As expected, the vast majority of well-characterized rocky planet masses fall below 10 M$_{\oplus}$, and the majority of planets more massive than this threshold are consistent with gas-rich compositions. We identify 4 planets with mass measurements exceeding 10 M$_{\oplus}$ and a radius consistent with a volatile-poor rocky composition, hereafter referred to as Mega-Earths: WASP-84 c, TOI-1347 b, TOI-2322 c, and HD 63433 b.  

We also consider GJ 523 b, despite sitting just outside the rocky planet density regime and therefore consistent with a modest gaseous envelope, as \cite{2026arXiv260324682K} identify this ultra-dense Neptune-sized object as a "Mega-Earth." Interior modeling suggests a Hycean (21$\%$ water mass) composition with little to no H/He envelope, assuming a standard four-layer planet composition (core, mantle, water layer, and H/He envelope) and an Earth-like ratio of iron to refractory abundances. Sitting near GJ 523 b in mass-radius space are three additional worlds worth considering: TOI-815 c, HD 18599 b, and GJ 143 b. Given that these worlds likely host little H/He envelope despite having sufficient mass to trigger runaway gas accretion we include them in this analysis. \footnote{HD 119130 b was originally found to have similar masses and radii, but \cite{2025AJ....169...26M} recently found this mass artificially inflated by a factor of three due to constructive interference with an untreated sinusoidal signal in the data such as stellar rotational modulation.} The eight planets identified as anomylously dense are highlighted in Figurer \ref{fig:massrad}.

%We include these four planets (GJ 523 b, GJ 413 b, TOI-815 c, and HD 18599 b) in our analysis as anomalously dense worlds, 

We quantify the distribution of planet masses and radii in our sample using a two-dimensional Gaussian Kernel Density Estimate (hereafter KDE) using \texttt{SciPy.stats}. To calculate the highest probability radii for any given value of planet mass p(R|M), we generated grids of planet radii from 0.001 to 4 R$_{\oplus}$ and planet mass from 0.001 to 50 M$_{\oplus}$ and then evaluated the KDE at each mass over the full grid of planetary radii. We then normalized this distribution over radius to get the conditional probability of radius for each slice in mass. The resulting probability density and our eight Mega-Earths are shown in Figure \ref{fig:massrad}. 

All eight Mega-Earths considered here correspond to probabilities $<$0.1, indicating that they have some of the most anomalous combinations of mass and radius of those in our sample. The planet corresponding to the lowest probability density is HD 63433 b (0.006), followed by HD 18599 b (0.007). 

\begin{figure*}
    \centering
    \includegraphics[width=0.9\linewidth]{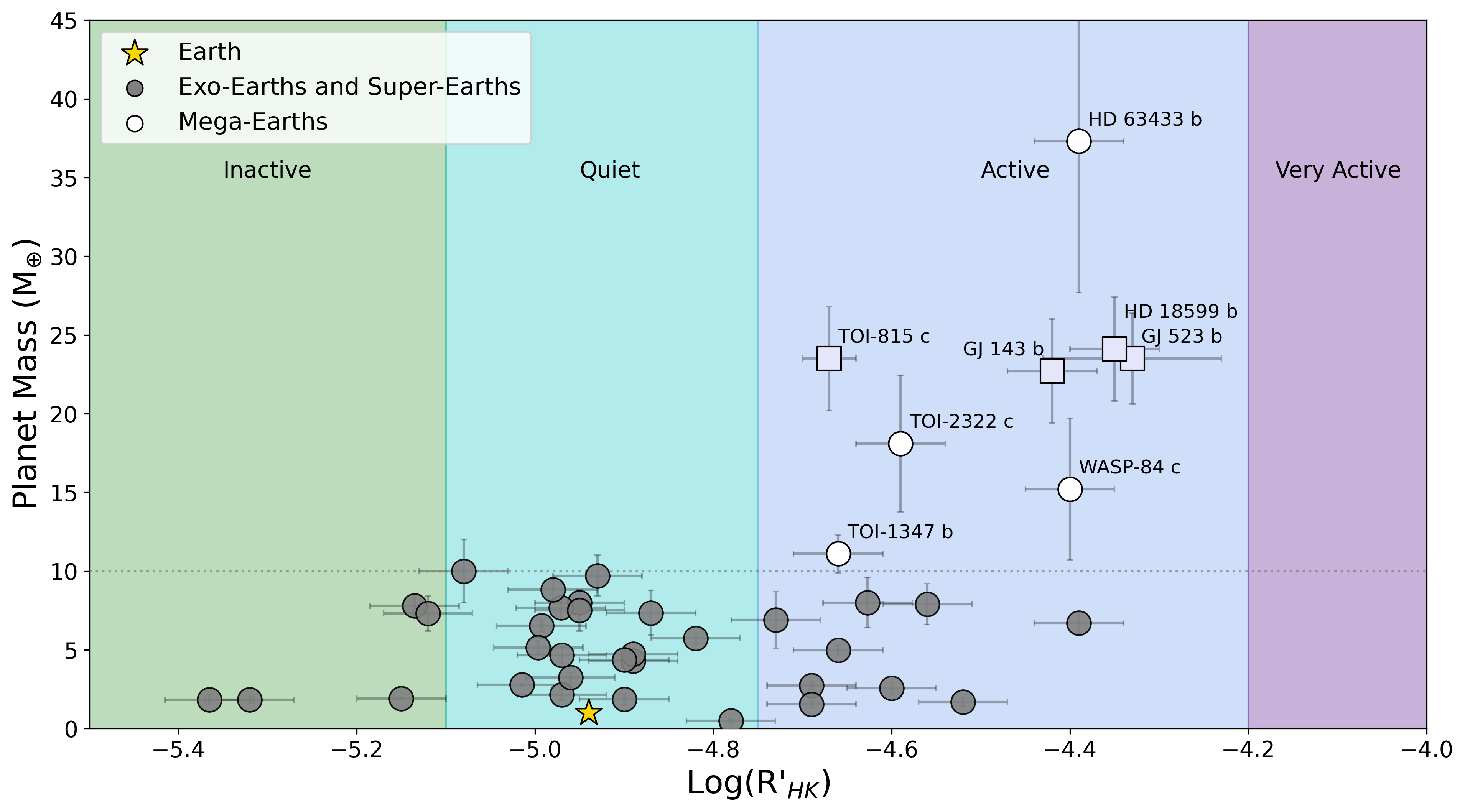}
    \caption{Planet Mass vs Stellar Activity indicators (here log$R'_{HK}$) for 37 planets consistent with rocky planet densities (circles). Planets with a mass below that of the runaway gas accretion threshhold (10 M$_{\oplus}$) are shown in grey while those above this limit are shown in white, delimited by a grey dotted line. Regions of high, very high, and low activity are highlighted \citep{1996AJ....111..439H}, with the Vaughn-Preston gap delimiting active and inactive stars. Planets consistent with volatile-rich compositions (GJ 523 b, GJ 143 b, TOI-815 c, and HD 18599 b) are shown with lavender squares to distinguish from those in the rocky planet density regime.}
    \label{fig:activity}
\end{figure*}

\subsection{WASP-84 c}
WASP-84 is 1--2 Gyr old K0 star hosting a transiting warm Jupiter and a recently discovered interior planet at an orbital period of $P=1.44$ days with a radius of $R_{c}=1.95 \pm 0.12 R_{\oplus}$ and a mass of $M_{c}=15.2 \pm 4.5 M_{\oplus}$ \citep{2023MNRAS.525L..43M}. The mass and radius measurements suggest an interior composition more iron-rich than the Earth, and the authors have dubbed this object a hot super-Earth. Complicating the RV analysis, WASP-84 exhibits chromospheric activity (log$R'_{\rm HK}=-4.42 \pm 0.04$) and starspots produce noticeable rotational modulation ($P_{\rm rot}=14.36$ days) in the TESS lightcurve (Fig \ref{fig:lightcurves}). Correspondingly, the authors report significant RV jitter of $\sim$13 m/s for the RV dataset, larger than the planetary signal ($K_{b}=7.7$ m/s).

%The radius measurement for this planet was obtained using two sectors of TESS photometry, while the mass was measured using 20 RVs from the CORALIE spectrograph and 22 additional RVs from HARPS \citep{2014MNRAS.445.1114A, 2015ApJ...800L...9A}. 

WASP-84 hosts not only a supposed Mega-Earth but also a warm Jupiter---neither of which likely formed in-situ. Interestingly, both planets have negligible eccentricity. Due to their well aligned orbits, both planets are thought to have migrated inwards through interactions with the proto-planetary disk, reducing the likelihood for chaotic post-disk-dispersal collisions \citep{2015ApJ...800L...9A}. 

\subsection{TOI-1347 b}

TOI-1347 is a 1-2 Gyr old, moderately active (log$R'_{\rm HK}=-4.66$) G-type star hosting an ultra-short period planet with a radius of $R_{c}=1.8 \pm 0.1 R_{\oplus}$ and a mass of $M_{c}=11.1 \pm 1.2 M_{\oplus}$ \citep{2024AJ....167..153R}. Lomb-scargle periodograms of both the photometry and RVs exhibit significant power at $\sim$8 days and at the likely stellar rotation period $P_{\rm rot}=16.34 \pm 0.57$ days. The orbital period of planet b, $P_{\rm orb}= 0.85$ days, is thankfully much shorter than the rotational period of the star. The RV semi-amplitude ($K_b=7.74$ m/s) measured using 134 RVs from the HIRES and HARPS-N spectrographs, however, is of similar order to the RV jitter ($\sigma$= 4.05 and 8.2 m/s respectively for HIRES and HARPS-N). 

The mass and radius measurements for TOI-1347 b suggest a bulk composition similar to that of Earth, and due to the mass and equilibrium temperature ($T_{\rm eq}=1400 \pm 40$ K), a H/He envelope can be ruled out with confidence. However, the phase curve in TESS photometry exhibits a secondary eclipse depth suggestive of heat transport through a gaseous envelope \citep{2024AJ....167..153R}, despite being sufficiently old for any primordial H/He envelope to have evaporated. This suggests the presence of a higher mean molecular weight envelope, necessitating the inner core to be even more dense than suggested by mass-radius curves. 

\subsection{TOI-2322 c}
TOI-2322 is a moderately active K-dwarf (logR'$_{\rm HK}=-4.59$) hosting two transiting rocky planets. While the inner rocky planet has a virtually unconstrained mass ($M_b=2.31 \pm 2.6 M_{\oplus}$), the outer has a radius of $R_{c}=1.87 \pm 0.66 R_{\oplus}$ and a mass of $M_{c}=18.1 \pm 4.34 M_{\oplus}$ \citep{hobson25}. The RV semi-amplitude of planet c ($K_c=5.37 \pm 1.6$ m/s), measured using 52 RVs from the ESPRESSO and HARPS-N spectrographs, is of similar scale to the RV jitter ($\sigma=7.4$ and $4.07$ m/s, respectively for ESPRESSO and HARPS-N). 

The rotational period of the star ($P_{\rm rot}=21.28 \pm 0.08$ days) is visible in the TESS light curves (Figure \ref{fig:lightcurves}), and disconcertingly coincides with the orbital period of TOI-2322 c ($P_{\rm orb_c}=20.23$ days). The orbital period of the inner  planet, TOI-2322 b, is roughly half this value ($P_{\rm orb_b}=11.31$ days). To attempt to disentangle these signals, \cite{hobson25} use a multivariate GP to simultaneously fit the photometry and RVs, as well as full-width half max and the bisector inverse slope activity indicators. 

\subsection{HD 63433 b}
HD 63433 is a young ($\sim$400 Myr) G-type star hosting three transiting planets \citep{2020AJ....160..179M, 2024AJ....167...54C}. HD 63433 b has the largest RV mass of any discussed here at $M_{b}=37.3 \pm 9.6 M_{\oplus}$ with a radius of $R_{b}=2.16 \pm 0.05 R_{\oplus}$  \citep{polanski2024}. This measurement was obtained by fitting 62 RVs from the HIRES and APF spectrographs, although notably \cite{mallorquin} do not measure a semi-amplitude for this planet using CARMENES RVs and report 3$\sigma$ upper limit of $M_b < 21 M_{\oplus}$. Both RV analyses use GPs informed by photometry. 

The planet semi-amplitude measured is $K_b=12.5 \pm 3.2$, which is similar in amplitude to the stellar jitter from HIRES ($\sigma=15$ m/s) and APF ($\sigma=5.0$). The rotational period of the host star ($P_{\rm rot}=6.45$ days) is close to the orbital period of HD 63433 b ($P_{\rm orb}=7.1$ days), complicating our ability to disentangle stellar and planetary RV signals. \cite{polanski2024} acknowledge this, predict that the GP is overfitting their RV dataset and producing an untrustworthy RV semi-amplitude, and caution putting too much faith into this mass measurement. For our analysis, we will use the \cite{polanski2024} mass value but harbor similar skepticism. 

\subsection{GJ 523 b}

With a mass of $M_b=23.5 \pm 3.3 M_{\oplus}$ and a radius of $R_b=2.55 \pm 0.15 R_{\oplus}$, GJ 523 b sits just above the rocky planet density regime in Figure \ref{fig:massrad}. This planet orbits a young (169 Myr), magnetically active star, though neither an S-value nor log$R'_{HK}$ are reported\footnote{We estimate this value using the the rotational period and convective turnover time in Section 3.1}. Rotational modulation with a periodicity of $P_{\rm rot}=5.621$ days is clearly visible in the TESS lightcurves (Figure \ref{fig:lightcurves}), and both the lightcurve and RVs show a peak in their respective LS periodograms at the rotational period of the star. To account for this activity, a Celerite Gaussian Process kernel was used to jointly fit the TESS lightcurve and NEID RVs, informed from the bisector inverse slope, which traces rotationally modulated activity in each RV measurement. The semi-amplitude of the planet ($K_b=6.87 \pm 0.95$ m/s) is larger than the stellar jitter.

GJ 523 b has an inclined orbit with respect to its host star($\psi > 0.71$) with relatively low eccentricity ($e=0.143$). Due to this unique combination of factors, \cite{2026arXiv260324682K} suggest GJ-523 b might have started as a primordial gas giant whose H/He envelope was lost during high eccentricity migration---however, the system is too young to have tidally circularized the planet's orbit. 

\subsection{TOI-815 c}  
The discovery paper for TOI-815 c reports a $R_c=2.94 \pm 0.15 R_{\oplus}$ planet with a mass of $M_c=23.5 \pm 2.5 M_{\oplus}$, along with a much less dense companion ($M_b=7.6 \pm 1.5 M_{\oplus}$, similar radius) using 34 RVs from VLT/ESPRESSO \citep{psaridi2024}. The semi-amplitude of planet c ($K_c=5.5 \pm 0.5$ m/s) is nearly three times the reported RV jitter (1.88 m/s). These planets orbit a young (0.2 Gyr) magnetically active K-dwarf with a reported median log$R'_{HK}=-4.67 \pm 0.03$. The rotation period of the star ($P_{\rm rot}=15.2 \pm 1.2$ days) is nearly half the orbital period of TOI-815 c ($P_{\rm orb}=34.9$ days), while the inner, lower density sub-Neptune has an orbital period of 11 days. To address the effects of stellar activity, \cite{psaridi2024} use a Matern 3/2 kernel Gaussian Process with photometrically informed priors and simultaneously fit the photometric and RV signals. 
%The authors suggest that the radically disparate densities of these two sub-Neptunes (with nearly identical radii) demonstrate the differences between warm and hot sub-Neptunes. 
%The authors report that this system has an inclination of $\sim 90 ^\circ$ and we are viewing the star nearly pole-on.

\subsection{HD 18599 b}
HD 18599 b is $R_b=2.60 \pm 0.15 R_{\oplus}$ planet with a reported mass of $M_b=24.1 \pm 7.1 M_{\oplus}$ orbiting a young (0.4 Gyr) magnetically active K-dwarf (log$R'_{HK}=-4.35 \pm 0.05$) \citep{desidera2023}. The authors utilized 112 RVs from HARPS and FEROS to measure a semi-amplitude of $K_b=11\pm 2$ m/s, which is of similar order to the reported RV jitter ($\sigma= 10 \pm 5$ m/s). %HD 18599 (also known as TOI-179) has a binary M-dwarf companion at close separation. 
The rotational period of HD 18599 ($P_{\rm rot}=8.73 \pm 0.7$ days) is roughly double the orbital period of planet b ($P_{\rm orb}=4.14$ days), and the authors use a quasi-perioid GP kernel to account for the effects of this activity on the RV signal. \cite{desidera2023} compare the high density of this planet to that of Kepler-411 b (with a TTV derived mass, \citep{2019A&A...624A..15S}), and to that of K2-233 c (a reported nearly pure-iron "super-Mercury", \citep{2020A&A...640A..48L}) that also orbit extremely young, active K-dwarfs. 

\subsection{GJ 143 b}
GJ 143 b has a radius of $R_b=2.61 \pm 0.17 R_{\oplus}$ and a mass of $M_b=22.7 \pm 2.2 M_{\oplus}$ \citep{2019ApJ...875L...7D} measured using a total of 140 RVs from HARPS and PFS. It orbits an active K-dwarf (log$R'_{HK}=-4.42 \pm 0.05$, \cite{boro2018}), with an orbital period nearly identical to the rotational period of its host star (35 days). The measured semi-amplitude ($K_b=5.51 \pm 0.41$ m/s) is larger than the reported RV jitter (1.44 m/s). The authors use a Gaussian Process (\texttt{celerite} kernel) fitting the RVs and photometry simultaneously to account for the effects of stellar activity. 

\section{Stellar Activity and Planet Mass}
\subsection{Mass vs log$R'_{HK}$}

\begin{table*}[ht]
\centering
\caption{Literature values for planet and host star properties. }
\label{tab:planet_properties}
\begin{tabular}{lcccccl}
\toprule
\textbf{Planet} &
\textbf{$M_p$ ($M_\oplus$)} &
\textbf{$R_p$ ($R_\oplus$)} &
\textbf{$P_{\rm orb}$ (days)} &
\textbf{$\log R'_{\rm HK}$} &
\textbf{References} \\
\midrule

\multicolumn{6}{l}{\textbf{Mega-Earths}} \\
\midrule

HD 63433 b
& $37.30 \pm 9.60$
& $2.16 \pm 0.06$
& $7.10794(7)$
& $-4.39 \pm 0.05$
& \cite{polanski2024, 2020AJ....160..179M} \\

TOI-1347 b
& $11.10 \pm 1.20$
& $1.80 \pm 0.10$
& $0.84742(6)$
& $-4.66 \pm 0.05$
& \cite{2024AJ....167..153R} \\

TOI-2322 c
& $18.10 \pm 4.34$
& $1.87 \pm 0.07$
& $20.2255(4)$
& $-4.59 \pm 0.05$
& \cite{hobson25}
\\

WASP-84 c
& $15.20 \pm 4.50$
& $1.95 \pm 0.12$
& $1.44689(2)$
& $-4.42 \pm 0.04$
& \cite{2023MNRAS.525L..43M} \\

GJ 523 b
& $23.5 \pm 3.3$
& $2.55 \pm 0.15$
& $7.10794(7)$
& $-4.3 \pm 0.1$*
& \cite{2026arXiv260324682K} \\

GJ 143 b
& $22.7 \pm 2.2$
& $2.61 \pm 0.17$
& $35.61253(6)$
& $-4.42 \pm 0.05$
& \cite{2019ApJ...875L...7D, boro2018} \\

TOI-815 c
& $23.5 \pm 2.5$
& $2.94 \pm 0.15$
& $11.197259(2)$
& $-4.67 \pm 0.03$
& \cite{psaridi2024} \\

HD 15899 b
& $24.1 \pm 7.1$
& $2.60 \pm 0.15$
& $4.1374354(4)$
& $-4.67 \pm 0.03$
& \cite{desidera2023} \\

\midrule
\multicolumn{6}{l}{\textbf{Normal Mass Rocky Planets}} \\
\midrule

55 Cnc e
& $7.99 \pm 0.32$
& $1.88 \pm 0.03$
& $0.73655(1)$
& $-5.04 \pm 0.04$
& \cite{bourrier2018, 2004ApJS..152..261W} 
\\

EPIC 201595106 b
& $7.67 \pm 0.75$
& $1.57 \pm 0.08$
& $0.87724(4)$
& $-4.97 \pm 0.03$
& \cite{2024NatSR..1427219L}
\\

EPIC 220674823 b
& $7.80 \pm 0.71$
& $1.68 \pm 0.04$
& $0.57131(6)$
& $-5.14\pm 0.1$
& \cite{2024MNRAS.529..141G, 2025ApJS..278...52H}
\\

GJ 1132 b
& $1.84 \pm 0.19$
& $1.19 \pm 0.04$
& $1.62893(3)$
& $-5.04\pm 0.1$
& \cite{2024ApJ...973L...8X, boro2018} 
\\

GJ 357 b
& $1.84 \pm 0.31$
& $1.20 \pm 0.06$
& $3.93060(2)$
& $-5.37\pm 0.1$
& \cite{weisserman26, boro2018} 
\\

GJ 367 b
& $0.50 \pm 0.08$
& $0.74 \pm 0.04$
& $0.321923(2)$
& $-4.74\pm 0.1$
& \cite{2026arXiv260618355L, boro2018}
\\

GJ 486 b
& $2.77 \pm 0.08$
& $1.29 \pm 0.02$
& $1.467121(3)$
& $-4.74\pm 0.1$
&  \cite{2024ApJ...975L..22W, boro2018}
\\

GJ 9827 b
& $4.28 \pm 0.35$
& $1.44 \pm 0.09$
& $1.20897(1)$
& $-5.28 \pm 0.05$
& \cite{passegger24}
\\

GJ 9827 c
& $1.86 \pm 0.37$
& $1.13 \pm 0.07$
& $3.64810(1)$
& $-5.28 \pm 0.05$
& \cite{passegger24}
\\

HD 15337 b
& $6.52 \pm 0.41$
& $1.77 \pm 0.03$
& $4.75598(6)$
& $-4.93 \pm 0.04$
&  \cite{rosario24, boro2018}
\\

HD 213885 b
& $8.83 \pm 0.66$
& $1.75 \pm 0.05$
& $1.00804(2)$
& $-4.90\pm 0.05$
& \cite{2020MNRAS.491.2982E, 1996AJ....111..439H}
\\

HD 219134 b
& $4.74 \pm 0.19$
& $1.60 \pm 0.06$
& $3.09293(1)$
& $-4.90\pm 0.1$
& \cite{2017NatAs...1E..56G, 2010ApJ...725..875I} 
\\

HD 219134 c
& $4.36 \pm 0.22$
& $1.51 \pm 0.05$
& $6.76458(3)$
& $-4.90\pm 0.1$
& \cite{2017NatAs...1E..56G, 2010ApJ...725..875I} 
\\

HD 260655 b
& $2.14 \pm 0.34$
& $1.24 \pm 0.02$
& $2.76953(3)$
& $-4.84 \pm 0.13$
& \cite{luque22} 
\\

HD 80653 b
& $5.72 \pm 0.36$
& $1.61 \pm 0.07$
& $0.71957(2)$
& $-4.82\pm 0.1$*
& \cite{bonomo23} 
\\

HIP 29442 d
& $5.14 \pm 0.41$
& $1.54 \pm 0.05$
& $6.42958(3)$
& $-5.00\pm 0.1$
&  \cite{egger24, boro2018}
\\

K2-131 b
& $7.90 \pm 1.30$
& $1.69 \pm 0.09$
& $0.36930(9)$
& $-4.56\pm 0.1$
& \cite{bonomo23, 2017AJ....154..226D} 
\\

K2-141 b
& $4.97 \pm 0.35$
& $1.51 \pm 0.05$
& $0.280324(2)$
& $-3.95 \pm 0.03$
& \cite{bonomo23, 2018AJ....155..107M}
\\

K2-199 b
& $6.90 \pm 1.80$
& $1.73 \pm 0.05$
& $3.22540(2)$
& $-4.74\pm 0.1$
& \cite{2021AJ....162..294A, 2025ApJS..278...52H} \\

K2-216 b
& $8.00 \pm 1.60$
& $1.75 \pm 0.17$
& $2.17480(5)$
& $-4.63\pm 0.1$
& \cite{persson18, 2025ApJS..278...52H}  
\\

K2-229 b
& $2.55 \pm 0.38$
& $1.26 \pm 0.08$
& $0.58422(2)$
& $-4.60\pm 0.1$
& \cite{2025ApJS..278...52H} \\

K2-265 b
& $7.34 \pm 1.43$
& $1.71 \pm 0.11$
& $2.36902(6)$
& $-4.87\pm 0.1$
& \cite{2023AJ....165..155T, 2025ApJS..278...52H} \\

K2-38 b
& $7.30 \pm 1.10$
& $1.54 \pm 0.14$
& $4.01593(5)$
& $-5.12\pm 0.1$
& \cite{toledo2020, 2025ApJS..278...52H} 
\\

Kepler-10 b
& $3.24 \pm 0.32$
& $1.47 \pm 0.03$
& $0.837491(2)$
& $-4.96 \pm 0.02$
& \cite{bonomo25, 2014ApJ...789..154D} 
\\

Kepler-107 c
& $10.00 \pm 2.00$
& $1.60 \pm 0.03$
& $4.901452(1)$
& $-5.08\pm 0.1$
& \cite{bonomo23, 2019NatAs...3..416B} 
\\

Kepler-20 b
& $9.70 \pm 1.30$
& $1.77 \pm 0.05$
& $3.69611(2)$
& $-4.93\pm 0.1$
& \cite{bonomo23, 2012ApJ...749...15G} 
\\

Kepler-21 b
& $7.50 \pm 1.30$
& $1.64 \pm 0.02$
& $2.78582(3)$
& $-4.95\pm 0.1$
& \cite{bonomo23, 2012ApJ...746..123H} 
\\

Kepler-78 b
& $1.68 \pm 0.27$
& $1.20 \pm 0.03$
& $0.355007(8)$
& $-4.52\pm 0.03$
& \cite{bonomo23, 2013Natur.503..381H} 
\\

Kepler-93 b
& $4.66 \pm 0.53$
& $1.48 \pm 0.02$
& $4.72674(1)$
& $-4.98\pm 0.07$
& \cite{bonomo23, claudi24} 
\\

LHS 1140 c
& $1.91 \pm 0.06$
& $1.27 \pm 0.03$
& $3.77794(2)$
& $-5.15\pm 0.1$
& \cite{2024ApJ...960L...3C, 2022MNRAS.514.2259S} \\

LTT 1445 A b
& $2.73 \pm 0.25$
& $1.34 \pm 0.11$
& $5.35876(4)$
& $-4.69\pm 0.1$
& \cite{2023AJ....166..171P, 2019AJ....158..152W} 
\\

LTT 1445 A c
& $1.54 \pm 0.20$
& $1.15 \pm 0.06$
& $3.12390(3)$
& $-4.69\pm 0.1$
& \cite{2023AJ....166..171P, 2019AJ....158..152W}  
\\

TOI-1235 b
& $6.69 \pm 0.67$
& $1.69 \pm 0.08$
& $3.44471(9)$
& $-4.39\pm 0.1$
& \cite{2022Sci...377.1211L, bluhm2020} 
\\

%TOI-500 b
%& $1.42 \pm 0.18$
%& $1.17 \pm 0.06$
%& $0.54817(2)$
%& $-4.67\pm 0.1$
%& \cite{2022NatAs...6..736S, }  
%\\

%WASP-47 e
%& $9.00 \pm 0.60$
%& $1.83 \pm 0.02$
%& $0.78961(3)$
%& $-5.20 \pm 0.1$
%& \cite{, 2025ApJS..278...52H}  \\

\bottomrule
\end{tabular}
\begin{tablenotes}
    \small
\item   Mass, radius, and orbital period for Mega-Earths are shown along with chromospheric activity of the host star. We've also provided values for the rocky planet sample for which we were able to obtain measurements for host star log$R'_{HK}$. For GJ 523 and other systems noted with a *, we estimate log$R'_{HK}$ using the method discussed in section 3.1. For published values of log$R'_{HK}$ without associated 1$\sigma$ uncertainties we have adopted a conservative value of 0.1. 
\end{tablenotes}

\end{table*}
All five of these Mega-Earths were identified as orbiting active stars in their respective discovery papers. We downloaded the TESS light curves for each target using the \texttt{Lightkurve} Python package \citep{2018ascl.soft12013L}. We removed NaN values, normalized the flux, binned the light curves by 35 cadences, and flattened the resulting light curves using a 1001-point window to remove long-term instrumental and systematic trends. We then removed $>6\sigma$ outliers and used the resulting light curves to visualize the rotational modulation of each host star (Figure \ref{fig:lightcurves}). 

In order to compare with the general population of normal-mass rocky planets, we identify literature values of log$R'_{HK}$ for 33 of the planets consistent with a rocky planet density in our dataset (Table \ref{tab:planet_properties}), including four of the five identified Mega-Earths. We use log$R'_{HK}$ as a tracer for magnetic activity in the host star, as the Ca II H $\&$ K lines are particularly sensitive to magnetic heating in the chromosphere, producing greater flux in the cores of these lines. 

For GJ 523, no value for log$R'_{HK}$ has been published---however, we can estimate the value using a typical Rossby number for a young ($\sim$160 Myr), rapidly rotating ($P_{\rm rot}=5.621$ d) K-dwarf and the relationship established in \cite{2008ApJ...687.1264M}. The Rossby number is defined as $R_{0}=\frac{P_{\rm rot}}{\tau_c}$, where $\tau_c$ is the convective turnover time. We estimate this timescale using B-V color magnitudes and Equation 4 in \cite{1984ApJ...287..769N}. We find a B-V of $\sim$1.0 \citep{2019AJ....158..138S}, which suggests a convective timescale of $\tau_c=23$ days. This would then correspond to a Rossby number of $\sim$0.25, and according to Fig. 7 and Equation 6 of \cite{2008ApJ...687.1264M} GJ 523 likely has a value of log$R'_{HK}= -4.33$. We adopt a conservative uncertainty estimate of 0.1 to account for systematics. 

We plot the literature values of planet mass as a function of log$R'_{HK}$ in Figure \ref{fig:activity}. We use the regions of log$R'_{HK}$ that were identified in \cite{1996AJ....111..439H} to determine if host stars are inactive, quiet, active, or very active, with log$R'_{HK}$=-4.75 corresponding to the boundary between quiet and active stars (Vaughn-Preston Gap) \citep{2009A&A...499L...9P}. While 24 stars fall in the inactive and quiet regions and 13 fall in the active region, all eight rocky and volatile-rich Mega-Earths orbit active stars with log$R'_{HK}>-4.75$.

If the underlying distribution of Mega-Earths across stellar activity levels was uniform, we would expect an equal likelihood of finding a mega-Earth around any star in our sample, regardless of activity. Considering first just the four rocky-density Mega-Earths, we can calculate the likelihood that they happen to orbit stars with log$R'_{HK}>-4.75$. The odds that a single Mega-Earth orbits an active star in our sample is 13/37, as we have mass and log$R'_{HK}$ values for 37 stars, 13 of which fall above the Vaughan-Preston gap. The odds that all four orbit active stars are:
\begin{center}
    p=$\frac{13}{37}\times\frac{12}{36}\times\frac{11}{35}\times\frac{10}{34}=0.01$
\end{center}

If we include GJ 523 b, GJ 413 b, TOI-815c, and HD 18599 b, this likelihood reduces to 0.0003. Given the slim likelihood that all mega-Earth discoveries happen to be around active stars, we are most likely biased toward finding these planets in such systems. This implies that either: 
\begin{enumerate}
    \item Young, active stars are more likely to host these extremely massive rocky planets.
    \item The effects of stellar activity the RVs for these systems has artificially inflated their mass measurements.
\end{enumerate}

\subsection{Should young stars preferentially host mega-Earths?}
Stellar chromospheric activity from calcium H and K lines is inversely associated with age \citep{1963ApJ...138..832W, 2008ApJ...687.1264M}. We expect certain categories of planets, such as inflated gas giants \citep{2019A&A...630A..81B} or those with primordial H/He envelopes \citep{2025MNRAS.539.2230R} to be more commonly observed around young stars. What about Mega-Earths?

The hypothesized pathways to produce dense rocky bodies more massive than 10 M$_{\oplus}$ require timescales consistent with, if not beyond that of traditional super-Earths. One method of circumventing runaway gas accretion is to form in a gas-free environment. For sun-like stars in clusters, roughly 80$\%$ are disk bearing at 1 Myr, 50-60$\%$ at 2-3 Myr, and only  5$\%$ at 10 Myr, with a typical disk lifetime of 5.8 $\pm$ 0.3 Myr \citep{2026A&A...707A.216P, 2016SSRv..205....3R}. While drag from the planet-forming disk helps to keep orbits circularized, many planets experience spikes in eccentricity as the disk disperses, triggering a period of giant impacts that turn planetesimals into planets. This period lasted $\sim$100 Myrs in the case of the solar system \citep{2015ApJ...810..136G, 2025ApJ...989..117K}. All rocky planets form from planetesimals during this phase--but to build a volatile-free planet with M$>$10 $_{\oplus}$, a larger than average number of planetesimal collisions are likely needed, requiring additional formation time, on average.

More problematic than the assembly timescale is the likelihood that these planets had to migrate inwards to their present day orbits. The Mega-Earths discussed here have orbits of 0.8--20 days, and the available solid inventory is almost certainly insufficient to form such massive planetary cores in situ \citep{2014ApJ...794...11I}. Smooth, low-eccentricity migration typically happens within the planet forming disk for massive planets that accreted rapidly \citep{2012A&A...546A..99K}, while higher eccentricity migration typically occurs post disk dispersal and requires subsequent tidal dissipation to produce circular orbits \citep{2018ARA&A..56..175D}, at least for giant planets. For rocky planets, planetary embryos likely migrated through the disk only to form planets post disk dispersal \citep{2014A&A...569A..56C}. The Mega-Earths in our sample all have nearly circular orbits, despite circularization timescales comparable to the stellar ages from a few Myr to several Gyr \citep{2018ARA&A..56..175D}. 

The second pathway to build a Mega-Earth is to form a giant planet with a gas-rich envelope and subsequently strip away the gas \citep{2020Natur.583...39A, 2019MNRAS.485.4454B}---a process which would also require the migration of a planet inwards. This stripping would almost certainly have to occur post disk-dispersal, given the high energy collisions needed to strip a giant planet's envelope and to avoid the resulting core re-accreting an envelope. Therefore high eccentricity migration would likely be necessary, requiring longer re-circularization times. 

The formation timescales of Mega-Earths are therefore at least as long as super-Earths, and there is no reason to believe that they would be likely to be destroyed or undergo a change in mass or radius through the lifetime of the planet. We therefore would not expect young stars to preferentially host these worlds, and predict at minimum a uniform distribution of Mega-Earths across host star age and activity level if not a preference toward older systems. 

\subsection{Are Mega-Earths a product of stellar activity?}

The remaining and most probable explanation for the fact that we observe several rocky planets with M$>$10M$_{\oplus}$ around active stars is that stellar activity has artificially inflated the RV mass measurements. Stellar activity introduces additional scatter to RV datasets, producing a fundamental noise floor \citep{2010ApJ...725..875I}--therefore we are less sensitive to planetary signals around an active star than we are to the planets with the same semi-amplitude around a magnetically quiet star of similar spectral type, size, and distance from Earth. When we couple this with our bias toward detecting and measuring larger semi-amplitudes over small ones that get swallowed up in the noise \citep{dumusque2011one, dumusque2011two}, as well as our bias toward publishing statistically significant planet detections with larger k/$\sigma_k$ ratios, this results in a population of planets with deceptively large masses orbiting only active stars. 

The predominant method used to fit the stellar variability---thereby allowing a more faithful interpretation of the planetary RV signals---is Gaussian Process (GP) Regression \citep{dumusque2011one}. GP regression describes the correlation between data through a covariance matrix and was introduced to the field of exoplanets in 2011 to model correlated noise from instrument systematics \citep{2012MNRAS.419.2683G}, and subsequently used to model correlated stellar variability due to the rotation and evolution of star spots \citep{2014MNRAS.443.2517H}. This method has been widely employed to produce mass measurements of small planets around active stars, including Kepler-78 b \citep{2016IAUFM..29A.208G} and K2-291 b \citep{2019AJ....157..116K}, amongst many more. 

However, our ability to disentangle these signals depends on the relative strengths of the RV semi-amplitude and stellar activity amplitude, the difference in the planet's orbital period and stellar rotation period, our ability to constrain these parameters and place meaningful priors on our GP, and the thoroughness of RV sampling and observing cadence. Overly flexible GPs can overfit an RV dataset, absorbing signal as stellar noise, often producing larger than expected planet semi-amplitudes and therefore masses \citep{2023AJ....166...62B}. 

The original analyses of all eight Mega-Earths above utilize GPs with various kernel forms (Celerite, Quasi-Periodic, 3/2 Matern, etc.), and all utilize prior constraints on rotational period and spot decay timescales informed by photometry. Despite taking all usual measures to ensure that the effects of stellar activity are accounted for in their RV analysis, the remaining mass measurements for these five planets remain improbably large. This calls into question our ability to fully account for the effects of stellar activity even when using well-constrained GPs. 

\cite{2017A&A...598A.133D} identify a critical threshold of signal to noise where GPs can reliably recover the underlying planetary RV signal from a noisy dataset. They find a critical threshold at $K/N=\frac{K_{pl}}{RV_{rms}}\sqrt{N_{obs}}=5$, but recommend a threshold of 7.5 for reliably detecting signals. Of the five Mega-Earths identified here, three fall below this threshold (WASP-84 c:  3.8, TOI-2322 c: 6.5, HD 63433 b: 5.3). HD 18599 b falls exactly at the 7.5 threshold, while the rest fall above with a range of 11-16. 

The systems where the orbital period of the planet---or harmonics thereof---is similar to the rotational period of the star are the most challenging to characterize, as any signal at these periods will be most enhanced by stellar activity. Of our Mega-Earth sample, GJ 143 b, HD 63433 b, and TOI-2322 c have an orbital period similar to the rotational period of their host star, while HD 18599 b and TOI-815 c orbit at harmonics of this value. \cite{polanski} acknowledge this challenge with HD 63433 b and caution putting faith in a RV-derived mass even with a well-informed GP. 

Despite having easily separable stellar rotation periods and planetary orbital periods, the remaining three planets are likely to have RV signals inflated by stellar activity. Looking forward, there is promise to better account for stellar activity using GPs when RVs are strategically scheduled to optimize the sampling of both magnetic activity and exoplanetary signals, that way both are more easily modeled \citep{2016AJ....152..204L, 2017A&A...608A..25B, 2020AJ....159...23N}. In particular, scheduling high-cadence RV measurements for short and ultra-short period planets with a nightly RV baseline offset could allow for a reduction in the effects of stellar activity on RV mass determination.

\section{Conclusions}
We analyze the population of planets with radii$<$4 R$_{\oplus}$ and find four worlds consistent with a rocky planet density that appear to be more massive than the gas-accretion threshold ($\sim$10 M$_{\oplus}$). We find four additional planets over this threshold that likely host dense volatile-rich atmospheres inconsistent with significant H/He content. These planets would require unusual formation pathways, such as orbital migration and envelope stripping or a series of larger than expected collisions post disk-dispersal. However, we find:

\begin{itemize}
    \item All eight Mega-Earths with M$>10 M_{\oplus}$ orbit active stars with log$R'_{HK}>-4.75$.
    \item There is no physical reason for Mega-Earths to prefer younger more active host stars---if anything, these worlds should prefer older systems to allow time for these unusal evolutionary pathways.
    \item We calculate a $<1\%$ likelihood that these mega-Earths coincidentally orbit exclusively active stars.
    \item The most likely culprit for these larger than expected masses is artificial inflation from stellar activity---particularly for systems that have similar stellar rotation and planetary orbital periods.
    \item Therefore, there is no compelling evidence for rocky planets forming above the upper mass limit of $10M_{\oplus}$.
\end{itemize} 

The mass measurements for these five worlds utilize photometrically informed Gaussian Processes to help disentangle stellar noise from planetary signal. Ultimately, despite our best efforts to characterize and fit stellar activity, we should remain skeptical of RV mass measurements for planets orbiting active stars---particularly so when the existence of such planets would challenge our notions of planet formation.

\section{Acknowledgements}
We would like to thank Andrew Cumming, Sabina Sagynbayeva, and Charles-Edouard Boukaré for helpful conversations. 

This work was supported by grant https://doi.org/10.69777/377645 from the Fonds de recherche du Québec and by the Center for research in astrophysics of Québec (AstroQuébec). The authors also thank the Trottier Space Institute (TSI) and l’Institut de recherche sur les exoplanètes (IREx) for their financial support and dynamic intellectual environment
%This work uses data obtained from multiple facilities at Maunakea Observatories. The authors wish to recognize the very significant cultural role and reverence that Maunakea has always had within the indigenous Hawaiian community, and the fact that the observatories are built on colonized sacred land. We recognize that her majesty Queen Lili‘uokalani yielded the Hawaiian Kingdom and these territories under duress and protest to the United States to avoid the bloodshed of her people. We further recognize that Hawai‘i remains an illegally occupied state of America. 

%This work is based on observations obtained at the international Gemini Observatory, a program of NSF NOIRLab, which is managed by the Association of Universities for Research in Astronomy (AURA) under a cooperative agreement with the U.S. National Science Foundation on behalf of the Gemini Observatory partnership: the U.S. National Science Foundation (United States), National Research Council (Canada), Agencia Nacional de Investigaci\'{o}n y Desarrollo (Chile), Ministerio de Ciencia, Tecnolog\'{i}a e Innovaci\'{o}n (Argentina), Minist\'{e}rio da Ci\^{e}ncia, Tecnologia, Inova\c{c}\~{o}es e Comunica\c{c}\~{o}es (Brazil), and Korea Astronomy and Space Science Institute (Republic of Korea).

%\facilities{Transiting Exoplanet Survey Satellite (TESS), W. M. Keck Observatory, Gemini Observatory}

\software{NumPy \citep{harris2020array}, Matplotlib \citep{Hunter:2007}, pandas \citep{mckinney-proc-scipy-2010}}, Lightkurve \citep{2018ascl.soft12013L}

\bibliography{paper}{}

@ARTICLE{1980PThPh..64..544M,
       author = {{Mizuno}, H.},
        title = "{Formation of the Giant Planets}",
      journal = {Progress of Theoretical Physics},
         year = 1980,
        month = aug,
       volume = {64},
       number = {2},
        pages = {544-557},
          doi = {10.1143/PTP.64.544},
       adsurl = {https://ui.adsabs.harvard.edu/abs/1980PThPh..64..544M}
}

@ARTICLE{2016SSRv..205....3R,
       author = {{Rab}, Christian and {Baldovin-Saavedra}, Carla and {Dionatos}, Odysseas and {Vorobyov}, Eduard and {G{\"u}del}, Manuel},
        title = "{The Gas Disk: Evolution and Chemistry}",
      journal = {\ssr},
         year = 2016,
        month = dec,
       volume = {205},
       number = {1-4},
        pages = {3-40},
          doi = {10.1007/s11214-016-0325-5},
archivePrefix = {arXiv},
       eprint = {1611.09658},
 primaryClass = {astro-ph.EP},
       adsurl = {https://ui.adsabs.harvard.edu/abs/2016SSRv..205....3R}
}

@ARTICLE{2026A&A...707A.216P,
       author = {{Polnitzky}, Fabian A. and {Ratzenb{\"o}ck}, Sebastian and {Gro{\ss}schedl}, Josefa and {Alves}, Jo{\~a}o},
        title = "{Precise determination of circumstellar disk lifetimes: Disk evolution in a single star-forming region}",
      journal = {\aap},
         year = 2026,
        month = mar,
       volume = {707},
          eid = {A216},
        pages = {A216},
          doi = {10.1051/0004-6361/202554921},
archivePrefix = {arXiv},
       eprint = {2512.06873},
 primaryClass = {astro-ph.SR},
       adsurl = {https://ui.adsabs.harvard.edu/abs/2026A&A...707A.216P}
}

@ARTICLE{1996Icar..124...62P,
       author = {{Pollack}, James B. and {Hubickyj}, Olenka and {Bodenheimer}, Peter and {Lissauer}, Jack J. and {Podolak}, Morris and {Greenzweig}, Yuval},
        title = "{Formation of the Giant Planets by Concurrent Accretion of Solids and Gas}",
      journal = {\icarus},
         year = 1996,
        month = nov,
       volume = {124},
       number = {1},
        pages = {62-85},
          doi = {10.1006/icar.1996.0190},
       adsurl = {https://ui.adsabs.harvard.edu/abs/1996Icar..124...62P}
}

@ARTICLE{2025ApJ...980L..40L,
       author = {{Lee}, Eve J. and {Owen}, James E.},
        title = "{Carving the Edges of the Rocky Planet Population}",
      journal = {\apjl},
         year = 2025,
        month = feb,
       volume = {980},
       number = {2},
          eid = {L40},
        pages = {L40},
          doi = {10.3847/2041-8213/adafa3},
archivePrefix = {arXiv},
       eprint = {2501.17241},
 primaryClass = {astro-ph.EP},
       adsurl = {https://ui.adsabs.harvard.edu/abs/2025ApJ...980L..40L}
}

@ARTICLE{2022MNRAS.517.3132R,
       author = {{Reinhardt}, Christian and {Meier}, Thomas and {Stadel}, Joachim G. and {Otegi}, Jon F. and {Helled}, Ravit},
        title = "{Forming iron-rich planets with giant impacts}",
      journal = {\mnras},
         year = 2022,
        month = dec,
       volume = {517},
       number = {3},
        pages = {3132-3143},
    
}

@ARTICLE{2016IAUFM..29A.208G,
       author = {{Grunblatt}, Samuel K. and {Howard}, Andrew W. and {Haywood}, Rapha{\"e}lle D.},
        title = "{Measuring the mass of Kepler-78b using nonparametric Gaussian process estimation}",
      journal = {IAU Focus Meeting},
         year = 2016,
        month = oct,
       volume = {29A},
        pages = {208-208},
          doi = {10.1017/S1743921316002829},
       adsurl = {https://ui.adsabs.harvard.edu/abs/2016IAUFM..29A.208G}
}

@ARTICLE{2019AJ....157..116K,
       author = {{Kosiarek}, Molly R. and {Blunt}, Sarah and {L{\'o}pez-Morales}, Mercedes and {Crossfield}, Ian J.~M. and {Sinukoff}, Evan and {Petigura}, Erik A. and {Gonzales}, Erica J. and {Poretti}, Ennio and {Malavolta}, Luca and {Howard}, Andrew W. and {Isaacson}, Howard and {Haywood}, Rapha{\"e}lle D. and {Ciardi}, David R. and {Bristow}, Makennah and {Collier Cameron}, Andrew and {Charbonneau}, David and {Dressing}, Courtney D. and {Figueira}, Pedro and {Fulton}, Benjamin J. and {Hardee}, Bronwen J. and {Hirsch}, Lea A. and {Latham}, David W. and {Mortier}, Annelies and {Nava}, Chantanelle and {Schlieder}, Joshua E. and {Vanderburg}, Andrew and {Weiss}, Lauren and {Bonomo}, Aldo S. and {Bouchy}, Fran{\c{c}}ois and {Buchhave}, Lars A. and {Coffinet}, Adrien and {Damasso}, Mario and {Dumusque}, Xavier and {Lovis}, Christophe and {Mayor}, Michel and {Micela}, Giusi and {Molinari}, Emilio and {Pepe}, Francesco and {Phillips}, David and {Piotto}, Giampaolo and {Rice}, Ken and {Sasselov}, Dimitar and {S{\'e}gransan}, Damien and {Sozzetti}, Alessandro and {Udry}, St{\'e}phane and {Watson}, Chris},
        title = "{K2-291b: A Rocky Super-Earth in a 2.2 day Orbit}",
      journal = {\aj},
         year = 2019,
        month = mar,
       volume = {157},
       number = {3},
          eid = {116},
        pages = {116},
          doi = {10.3847/1538-3881/aafe83},
archivePrefix = {arXiv},
       eprint = {1901.04558},
 primaryClass = {astro-ph.EP},
       adsurl = {https://ui.adsabs.harvard.edu/abs/2019AJ....157..116K}
}

@ARTICLE{2024ApJ...960L...3C,
       author = {{Cadieux}, Charles and {Plotnykov}, Mykhaylo and {Doyon}, Ren{\'e} and {Valencia}, Diana and {Jahandar}, Farbod and {Dang}, Lisa and {Turbet}, Martin and {Fauchez}, Thomas J. and {Cloutier}, Ryan and {Cherubim}, Collin and {Artigau}, {\'E}tienne and {Cook}, Neil J. and {Edwards}, Billy and {Hallatt}, Tim and {Charnay}, Benjamin and {Bouchy}, Fran{\c{c}}ois and {Allart}, Romain and {Mignon}, Lucile and {Baron}, Fr{\'e}d{\'e}rique and {Barros}, Susana C.~C. and {Benneke}, Bj{\"o}rn and {Canto Martins}, B.~L. and {Cowan}, Nicolas B. and {De Medeiros}, J.~R. and {Delfosse}, Xavier and {Delgado-Mena}, Elisa and {Dumusque}, Xavier and {Ehrenreich}, David and {Frensch}, Yolanda G.~C. and {Gonz{\'a}lez Hern{\'a}ndez}, J.~I. and {Hara}, Nathan C. and {Lafreni{\`e}re}, David and {Lo Curto}, Gaspare and {Malo}, Lison and {Melo}, Claudio and {Mounzer}, Dany and {Passeger}, Vera Maria and {Pepe}, Francesco and {Poulin-Girard}, Anne-Sophie and {Santos}, Nuno C. and {Sosnowska}, Danuta and {Su{\'a}rez Mascare{\~n}o}, Alejandro and {Thibault}, Simon and {Vaulato}, Valentina and {Wade}, Gregg A. and {Wildi}, Fran{\c{c}}ois},
        title = "{New Mass and Radius Constraints on the LHS 1140 Planets: LHS 1140 b Is either a Temperate Mini-Neptune or a Water World}",
      journal = {\apjl},
         year = 2024,
        month = jan,
       volume = {960},
       number = {1},
          eid = {L3},
        pages = {L3},
          doi = {10.3847/2041-8213/ad1691},
archivePrefix = {arXiv},
       eprint = {2310.15490},
 primaryClass = {astro-ph.EP},
       adsurl = {https://ui.adsabs.harvard.edu/abs/2024ApJ...960L...3C}
}

@ARTICLE{2012ApJ...746..123H,
       author = {{Howell}, Steve B. and {Rowe}, Jason F. and {Bryson}, Stephen T. and {Quinn}, Samuel N. and {Marcy}, Geoffrey W. and {Isaacson}, Howard and {Ciardi}, David R. and {Chaplin}, William J. and {Metcalfe}, Travis S. and {Monteiro}, Mario J.~P.~F.~G. and {Appourchaux}, Thierry and {Basu}, Sarbani and {Creevey}, Orlagh L. and {Gilliland}, Ronald L. and {Quirion}, Pierre-Olivier and {Stello}, Denis and {Kjeldsen}, Hans and {Christensen-Dalsgaard}, J{\"o}rgen and {Elsworth}, Yvonne and {Garc{\'\i}a}, Rafael A. and {Houdek}, G{\"u}nter and {Karoff}, Christoffer and {Molenda-{\.Z}akowicz}, Joanna and {Thompson}, Michael J. and {Verner}, Graham A. and {Torres}, Guillermo and {Fressin}, Francois and {Crepp}, Justin R. and {Adams}, Elisabeth and {Dupree}, Andrea and {Sasselov}, Dimitar D. and {Dressing}, Courtney D. and {Borucki}, William J. and {Koch}, David G. and {Lissauer}, Jack J. and {Latham}, David W. and {Buchhave}, Lars A. and {Gautier}, III, Thomas N. and {Everett}, Mark and {Horch}, Elliott and {Batalha}, Natalie M. and {Dunham}, Edward W. and {Szkody}, Paula and {Silva}, David R. and {Mighell}, Ken and {Holberg}, Jay and {Ballot}, Jer{\^o}me and {Bedding}, Timothy R. and {Bruntt}, Hans and {Campante}, Tiago L. and {Handberg}, Rasmus and {Hekker}, Saskia and {Huber}, Daniel and {Mathur}, Savita and {Mosser}, Benoit and {R{\'e}gulo}, Clara and {White}, Timothy R. and {Christiansen}, Jessie L. and {Middour}, Christopher K. and {Haas}, Michael R. and {Hall}, Jennifer R. and {Jenkins}, Jon M. and {McCaulif}, Sean and {Fanelli}, Michael N. and {Kulesa}, Craig and {McCarthy}, Don and {Henze}, Christopher E.},
        title = "{Kepler-21b: A 1.6 R $_{Earth}$ Planet Transiting the Bright Oscillating F Subgiant Star HD 179070}",
      journal = {\apj},
         year = 2012,
        month = feb,
       volume = {746},
       number = {2},
          eid = {123},
        pages = {123},
          doi = {10.1088/0004-637X/746/2/123},
archivePrefix = {arXiv},
       eprint = {1112.2165},
 primaryClass = {astro-ph.SR},
       adsurl = {https://ui.adsabs.harvard.edu/abs/2012ApJ...746..123H}
}

@ARTICLE{bonomo25,
       author = {{Bonomo}, A.~S. and {Borsato}, L. and {Rajpaul}, V.~M. and {Zeng}, L. and {Damasso}, M. and {Hara}, N.~C. and {Cretignier}, M. and {Leleu}, A. and {Unger}, N. and {Dumusque}, X. and {Lienhard}, F. and {Mortier}, A. and {Naponiello}, L. and {Malavolta}, L. and {Sozzetti}, A. and {Latham}, D.~W. and {Rice}, K. and {Bongiolatti}, R. and {Buchhave}, L. and {Cameron}, A.~C. and {Fiorenzano}, A.~F. and {Ghedina}, A. and {Haywood}, R.~D. and {Lacedelli}, G. and {Massa}, A. and {Pepe}, F. and {Poretti}, E. and {Udry}, S.},
        title = "{In-depth characterization of the Kepler-10 three-planet system with HARPS-N radial velocities and Kepler transit timing variations}",
      journal = {\aap},
         year = 2025,
        month = apr,
       volume = {696},
          eid = {A233},
        pages = {A233},
          doi = {10.1051/0004-6361/202453026},
archivePrefix = {arXiv},
       eprint = {2502.07996},
 primaryClass = {astro-ph.EP},
       adsurl = {https://ui.adsabs.harvard.edu/abs/2025A&A...696A.233B}
}

@ARTICLE{2014MNRAS.443.2517H,
       author = {{Haywood}, R.~D. and {Collier Cameron}, A. and {Queloz}, D. and {Barros}, S.~C.~C. and {Deleuil}, M. and {Fares}, R. and {Gillon}, M. and {Lanza}, A.~F. and {Lovis}, C. and {Moutou}, C. and {Pepe}, F. and {Pollacco}, D. and {Santerne}, A. and {S{\'e}gransan}, D. and {Unruh}, Y.~C.},
        title = "{Planets and stellar activity: hide and seek in the CoRoT-7 system}",
      journal = {\mnras},
         year = 2014,
        month = sep,
       volume = {443},
       number = {3},
        pages = {2517-2531},
          doi = {10.1093/mnras/stu1320},
archivePrefix = {arXiv},
       eprint = {1407.1044},
 primaryClass = {astro-ph.EP},
       adsurl = {https://ui.adsabs.harvard.edu/abs/2014MNRAS.443.2517H}
}

@ARTICLE{2020A&A...634A..43O,
       author = {{Otegi}, J.~F. and {Bouchy}, F. and {Helled}, R.},
        title = "{Revisited mass-radius relations for exoplanets below 120 M$_{{\ensuremath{\oplus}}}$}",
      journal = {\aap},
         year = 2020,
        month = feb,
       volume = {634},
          eid = {A43},
        pages = {A43},
          doi = {10.1051/0004-6361/201936482},
archivePrefix = {arXiv},
       eprint = {1911.04745},
 primaryClass = {astro-ph.EP},
       adsurl = {https://ui.adsabs.harvard.edu/abs/2020A&A...634A..43O}
}

@ARTICLE{2017A&A...598A.133D,
       author = {{Dumusque}, X. and {Borsa}, F. and {Damasso}, M. and {D{\'\i}az}, R.~F. and {Gregory}, P.~C. and {Hara}, N.~C. and {Hatzes}, A. and {Rajpaul}, V. and {Tuomi}, M. and {Aigrain}, S. and {Anglada-Escud{\'e}}, G. and {Bonomo}, A.~S. and {Bou{\'e}}, G. and {Dauvergne}, F. and {Frustagli}, G. and {Giacobbe}, P. and {Haywood}, R.~D. and {Jones}, H.~R.~A. and {Laskar}, J. and {Pinamonti}, M. and {Poretti}, E. and {Rainer}, M. and {S{\'e}gransan}, D. and {Sozzetti}, A. and {Udry}, S.},
        title = "{Radial-velocity fitting challenge. II. First results of the analysis of the data set}",
      journal = {\aap},
         year = 2017,
        month = feb,
       volume = {598},
          eid = {A133},
        pages = {A133},
          doi = {10.1051/0004-6361/201628671},
archivePrefix = {arXiv},
       eprint = {1609.03674},
 primaryClass = {astro-ph.EP},
       adsurl = {https://ui.adsabs.harvard.edu/abs/2017A&A...598A.133D}
}

@ARTICLE{2023AJ....166...62B,
       author = {{Blunt}, Sarah and {Carvalho}, Adolfo and {David}, Trevor J. and {Beichman}, Charles and {Zink}, Jon K. and {Gaidos}, Eric and {Behmard}, Aida and {Bouma}, Luke G. and {Cody}, Devin and {Dai}, Fei and {Foreman-Mackey}, Daniel and {Grunblatt}, Sam and {Howard}, Andrew W. and {Kosiarek}, Molly and {Knutson}, Heather A. and {Rubenzahl}, Ryan A. and {Beard}, Corey and {Chontos}, Ashley and {Giacalone}, Steven and {Hirano}, Teruyuki and {Johnson}, Marshall C. and {Lubin}, Jack and {Akana Murphy}, Joseph M. and {Petigura}, Erik A. and {Van Zandt}, Judah and {Weiss}, Lauren},
        title = "{Overfitting Affects the Reliability of Radial Velocity Mass Estimates of the V1298 Tau Planets}",
      journal = {\aj},
         year = 2023,
        month = aug,
       volume = {166},
       number = {2},
          eid = {62},
        pages = {62},
          doi = {10.3847/1538-3881/acde78},
archivePrefix = {arXiv},
       eprint = {2306.08145},
 primaryClass = {astro-ph.EP},
       adsurl = {https://ui.adsabs.harvard.edu/abs/2023AJ....166...62B}
}

@ARTICLE{2025AJ....169...26M,
       author = {{Akana Murphy}, Joseph M. and {Luque}, Rafael and {Batalha}, Natalie M. and {Carleo}, Ilaria and {Palle}, Enric and {Brady}, Madison and {Fulton}, Benjamin and {Handley}, Luke B. and {Isaacson}, Howard and {Lacedelli}, Gaia and {Murgas}, Felipe and {Nowak}, Grzegorz and {Orell-Miquel}, J. and {Osborne}, Hannah L.~M. and {Van Eylen}, Vincent and {Zapatero Osorio}, Mar{\'\i}a Rosa},
        title = "{HD 119130 b Is Not an ``Ultradense'' Sub-Neptune}",
      journal = {\aj},
         year = 2025,
        month = jan,
       volume = {169},
       number = {1},
          eid = {26},
        pages = {26},
          doi = {10.3847/1538-3881/ad89c0},
archivePrefix = {arXiv},
       eprint = {2411.02518},
 primaryClass = {astro-ph.EP},
       adsurl = {https://ui.adsabs.harvard.edu/abs/2025AJ....169...26M}
}

@ARTICLE{2019A&A...624A..15S,
       author = {{Sun}, L. and {Ioannidis}, P. and {Gu}, S. and {Schmitt}, J.~H.~M.~M. and {Wang}, X. and {Kouwenhoven}, M.~B.~N.},
        title = "{Kepler-411: a four-planet system with an active host star}",
      journal = {\aap},
         year = 2019,
        month = apr,
       volume = {624},
          eid = {A15},
        pages = {A15},
          doi = {10.1051/0004-6361/201834275},
archivePrefix = {arXiv},
       eprint = {1902.09719},
 primaryClass = {astro-ph.EP},
       adsurl = {https://ui.adsabs.harvard.edu/abs/2019A&A...624A..15S}
}

@ARTICLE{2020A&A...640A..48L,
       author = {{Lillo-Box}, J. and {Lopez}, T.~A. and {Santerne}, A. and {Nielsen}, L.~D. and {Barros}, S.~C.~C. and {Deleuil}, M. and {Acu{\~n}a}, L. and {Mousis}, O. and {Sousa}, S.~G. and {Adibekyan}, V. and {Armstrong}, D.~J. and {Barrado}, D. and {Bayliss}, D. and {Brown}, D.~J.~A. and {Demangeon}, O.~D.~S. and {Dumusque}, X. and {Figueira}, P. and {Hojjatpanah}, S. and {Osborn}, H.~P. and {Santos}, N.~C. and {Udry}, S.},
        title = "{Masses for the seven planets in K2-32 and K2-233. Four diverse planets in resonant chain and the first young rocky worlds}",
      journal = {\aap},
         year = 2020,
        month = aug,
       volume = {640},
          eid = {A48},
        pages = {A48},
          doi = {10.1051/0004-6361/202037896},
archivePrefix = {arXiv},
       eprint = {2006.01102},
 primaryClass = {astro-ph.EP},
       adsurl = {https://ui.adsabs.harvard.edu/abs/2020A&A...640A..48L}
}

@ARTICLE{desidera2023,
       author = {{Desidera}, S. and {Damasso}, M. and {Gratton}, R. and {Benatti}, S. and {Nardiello}, D. and {D'Orazi}, V. and {Lanza}, A.~F. and {Locci}, D. and {Marzari}, F. and {Mesa}, D. and {Messina}, S. and {Pillitteri}, I. and {Sozzetti}, A. and {Girard}, J. and {Maggio}, A. and {Micela}, G. and {Malavolta}, L. and {Nascimbeni}, V. and {Pinamonti}, M. and {Squicciarini}, V. and {Alcal{\'a}}, J. and {Biazzo}, K. and {Bohn}, A. and {Bonavita}, M. and {Brooks}, K. and {Chauvin}, G. and {Covino}, E. and {Delorme}, P. and {Hagelberg}, J. and {Janson}, M. and {Lagrange}, A.-M. and {Lazzoni}, C.},
        title = "{TOI-179: A young system with a transiting compact Neptune-mass planet and a low-mass companion in outer orbit}",
      journal = {\aap},
         year = 2023,
        month = jul,
       volume = {675},
          eid = {A158},
        pages = {A158},
          doi = {10.1051/0004-6361/202244611},
archivePrefix = {arXiv},
       eprint = {2210.07933},
 primaryClass = {astro-ph.EP},
       adsurl = {https://ui.adsabs.harvard.edu/abs/2023A&A...675A.158D}
}

@ARTICLE{psaridi2024,
       author = {{Psaridi}, Angelica and {Osborn}, Hugh and {Bouchy}, Fran{\c{c}}ois and {Lendl}, Monika and {Parc}, L{\'e}na and {Billot}, Nicolas and {Broeg}, Christopher and {Sousa}, S{\'e}rgio G. and {Adibekyan}, Vardan and {Attia}, Mara and {Bonfanti}, Andrea and {Chakraborty}, Hritam and {Collins}, Karen A. and {Davoult}, Jeanne and {Delgado-Mena}, Elisa and {Grieves}, Nolan and {Guillot}, Tristan and {Heitzmann}, Alexis and {Helled}, Ravit and {Hellier}, Coel and {Jenkins}, Jon M. and {Knierim}, Henrik and {Krenn}, Andreas and {Lissauer}, Jack J. and {Luque}, Rafael and {Rapetti}, David and {Santos}, Nuno C. and {Su{\'a}rez}, Olga and {Venturini}, Julia and {Wilkin}, Francis P. and {Wilson}, Thomas G. and {Winn}, Joshua N. and {Ziegler}, Carl and {Zingales}, Tiziano and {Alibert}, Yann and {Brandeker}, Alexis and {Egger}, Jo Ann and {Gandolfi}, Davide and {Hooton}, Matthew J. and {Tuson}, Amy and {Ulmer-Moll}, Sol{\`e}ne and {Abe}, Lyu and {Allart}, Romain and {Alonso}, Roi and {Anderson}, David R. and {Escud{\'e}}, Guillem Anglada and {B{\'a}rczy}, Tamas and {Barrado}, David and {Barros}, Susana C.~C. and {Baumjohann}, Wolfgang and {Beck}, Mathias and {Beck}, Thomas and {Benz}, Willy and {Bonfils}, Xavier and {Borsato}, Luca and {Bourrier}, Vincent and {Ciardi}, David R. and {Cameron}, Andrew Collier and {Charnoz}, S{\'e}bastien and {Cointepas}, Marion and {Csizmadia}, Szil{\'a}rd and {Cubillos}, Patricio and {Lo Curto}, Gaspare and {Davies}, Melvyn B. and {Daylan}, Tansu and {Deleuil}, Magali and {Deline}, Adrien and {Delrez}, Laetitia and {Demangeon}, Olivier D.~S. and {Demory}, Brice-Olivier and {Dorn}, Caroline and {Dumusque}, Xavier and {Ehrenreich}, David and {Erikson}, Anders and {Lecavelier des Etangs}, Alain and {de Miguel}, Diana and {Fortier}, Andrea and {Fossati}, Luca and {Frensch}, Yolanda G.~C. and {Fridlund}, Malcolm and {Gillon}, Micha{\"e}l and {G{\"u}del}, Manuel and {G{\"u}nther}, Maximilian N. and {Hagelberg}, Janis and {Helling}, Christiane and {Hoyer}, Sergio and {Isaak}, Kate G. and {Kiss}, Laszlo L. and {Lam}, Kristine W.~F. and {Laskar}, Jacques and {Lavie}, Baptiste and {Lovis}, Christophe and {Magrin}, Demetrio and {Marafatto}, Luca and {Maxted}, Pierre and {McDermott}, Scott and {M{\'e}karnia}, Djamel and {Mordasini}, Christoph and {Murgas}, Felipe and {Nascimbeni}, Valerio and {Nielsen}, Louise D. and {Olofsson}, G{\"o}ran and {Ottensamer}, Roland and {Pagano}, Isabella and {Pall{\'e}}, Enric and {Peter}, Gisbert and {Piotto}, Giampaolo and {Pollacco}, Don and {Queloz}, Didier and {Ragazzoni}, Roberto and {Ramos}, Devin and {Rando}, Nicola and {Rauer}, Heike and {Reimers}, Christian and {Ribas}, Ignasi and {Seager}, Sara and {S{\'e}gransan}, Damien and {Scandariato}, Gaetano and {Simon}, Attila E. and {Smith}, Alexis M.~S. and {Stalport}, Manu and {Steller}, Manfred and {Szab{\'o}}, Gyula and {Thomas}, Nicolas and {Pritchard}, Tyler A. and {Udry}, St{\'e}phane and {Corral Van Damme}, Carlos and {Van Grootel}, Val{\'e}rie and {Villaver}, Eva and {Walter}, Ingo and {Walton}, Nicholas and {Watkins}, Cristilyn N. and {West}, Richard G.},
        title = "{Discovery of two warm mini-Neptunes with contrasting densities orbiting the young K3V star TOI-815}",
      journal = {\aap},
         year = 2024,
        month = may,
       volume = {685},
          eid = {A5},
        pages = {A5},
          doi = {10.1051/0004-6361/202348678},
archivePrefix = {arXiv},
       eprint = {2401.15709},
 primaryClass = {astro-ph.EP},
       adsurl = {https://ui.adsabs.harvard.edu/abs/2024A&A...685A...5P}
}

@ARTICLE{hobson25,
       author = {{Hobson}, M.~J. and {Su{\'a}rez Mascare{\~n}o}, A. and {Lovis}, C. and {Bouchy}, F. and {Lavie}, B. and {Cretignier}, M. and {Silva}, A.~M. and {Sousa}, S.~G. and {Tabernero}, H.~M. and {Adibekyan}, V. and {Allende Prieto}, C. and {Alibert}, Y. and {Barros}, S.~C.~C. and {Castro-Gonz{\'a}lez}, A. and {Collins}, K.~A. and {Cristiani}, S. and {D'Odorico}, V. and {Damasso}, M. and {Dragomir}, D. and {Dumusque}, X. and {Ehrenreich}, D. and {Figueira}, P. and {G{\'e}nova Santos}, R. and {Goeke}, B. and {Gonz{\'a}lez Hern{\'a}ndez}, J.~I. and {Hesse}, K. and {Lillo-Box}, J. and {Lo Curto}, G. and {Martins}, C.~J.~A.~P. and {Mehner}, A. and {Micela}, G. and {Molaro}, P. and {Nunes}, N.~J. and {Palle}, E. and {Passegger}, V.~M. and {Pepe}, F. and {Rebolo}, R. and {Rodrigues}, J. and {Santos}, N. and {Sozzetti}, A. and {Tofflemire}, B.~M. and {Udry}, S. and {Watkins}, C. and {Zapatero Osorio}, M.-R. and {Ziegler}, C.},
        title = "{TOI-2322: Two transiting rocky planets close to the stellar rotation period and its first harmonic}",
      journal = {\aap},
         year = 2025,
        month = oct,
       volume = {702},
          eid = {A32},
        pages = {A32},
          doi = {10.1051/0004-6361/202555614},
archivePrefix = {arXiv},
       eprint = {2508.18094},
 primaryClass = {astro-ph.EP},
       adsurl = {https://ui.adsabs.harvard.edu/abs/2025A&A...702A..32H}
}

@ARTICLE{2024AJ....167..153R,
       author = {{Rubenzahl}, Ryan A. and {Dai}, Fei and {Howard}, Andrew W. and {Lissauer}, Jack J. and {Van Zandt}, Judah and {Beard}, Corey and {Giacalone}, Steven and {Murphy}, Joseph M. Akana and {Chontos}, Ashley and {Lubin}, Jack and {Brinkman}, Casey L. and {Tyler}, Dakotah and {MacDougall}, Mason G. and {Rice}, Malena and {Dalba}, Paul A. and {Mayo}, Andrew W. and {Weiss}, Lauren M. and {Polanski}, Alex S. and {Blunt}, Sarah and {Yee}, Samuel W. and {Hill}, Michelle L. and {Angelo}, Isabel and {Turtelboom}, Emma V. and {Holcomb}, Rae and {Behmard}, Aida and {Pidhorodetska}, Daria and {Batalha}, Natalie M. and {Crossfield}, Ian J.~M. and {Dressing}, Courtney and {Fulton}, Benjamin and {Huber}, Daniel and {Isaacson}, Howard and {Kane}, Stephen R. and {Petigura}, Erik A. and {Robertson}, Paul and {Scarsdale}, Nicholas and {Mocnik}, Teo and {Fetherolf}, Tara and {Malavolta}, Luca and {Mortier}, Annelies and {Fiorenzano}, Aldo and {Pedani}, Marco},
        title = "{The TESS-Keck Survey. XII. A Dense 1.8 R $_{{\ensuremath{\oplus}}}$ Ultra-short-period Planet Possibly Clinging to a High-mean-molecular-weight Atmosphere after the First Gigayear}",
      journal = {\aj},
         year = 2024,
        month = apr,
       volume = {167},
       number = {4},
          eid = {153},
        pages = {153},
          doi = {10.3847/1538-3881/ad28bb},
archivePrefix = {arXiv},
       eprint = {2402.07451},
 primaryClass = {astro-ph.EP},
       adsurl = {https://ui.adsabs.harvard.edu/abs/2024AJ....167..153R}
}

@ARTICLE{polanski,
       author = {{Polanski}, Alex S. and {Lubin}, Jack and {Beard}, Corey and {Akana Murphy}, Joseph M. and {Rubenzahl}, Ryan and {Hill}, Michelle L. and {Crossfield}, Ian J.~M. and {Chontos}, Ashley and {Robertson}, Paul and {Isaacson}, Howard and {Kane}, Stephen R. and {Ciardi}, David R. and {Batalha}, Natalie M. and {Dressing}, Courtney and {Fulton}, Benjamin and {Howard}, Andrew W. and {Huber}, Daniel and {Petigura}, Erik A. and {Weiss}, Lauren M. and {Angelo}, Isabel and {Behmard}, Aida and {Blunt}, Sarah and {Brinkman}, Casey L. and {Dai}, Fei and {Dalba}, Paul A. and {Fetherolf}, Tara and {Giacalone}, Steven and {Hirsch}, Lea A. and {Holcomb}, Rae and {Kosiarek}, Molly R. and {Mayo}, Andrew W. and {MacDougall}, Mason G. and {Mo{\v{c}}nik}, Teo and {Pidhorodetska}, Daria and {Rice}, Malena and {Rosenthal}, Lee J. and {Scarsdale}, Nicholas and {Turtelboom}, Emma V. and {Tyler}, Dakotah and {Van Zandt}, Judah and {Yee}, Samuel W. and {Coria}, David R. and {Dulz}, Shannon D. and {Hartman}, Joel D. and {Householder}, Aaron and {Lange}, Sarah and {Langford}, Andrew and {Louden}, Emma M. and {Siegel}, Jared C. and {Gilbert}, Emily A. and {Gonzales}, Erica J. and {Schlieder}, Joshua E. and {Boyle}, Andrew W. and {Christiansen}, Jessie L. and {Clark}, Catherine A. and {Fernandes}, Rachel B. and {Lund}, Michael B. and {Savel}, Arjun B. and {Gill}, Holden and {Beichman}, Charles and {Matson}, Rachel and {Matthews}, Elisabeth C. and {Furlan}, E. and {Howell}, Steve B. and {Scott}, Nicholas J. and {Everett}, Mark E. and {Livingston}, John H. and {Ershova}, Irina O. and {Cheryasov}, Dmitry V. and {Safonov}, Boris and {Lillo-Box}, Jorge and {Barrado}, David and {Morales-Calder{\'o}n}, Mar{\'\i}a},
        title = "{The TESS-Keck Survey. XX. 15 New TESS Planets and a Uniform RV Analysis of All Survey Targets}",
      journal = {\apjs},
         year = 2024,
        month = jun,
       volume = {272},
       number = {2},
          eid = {32},
        pages = {32},
          doi = {10.3847/1538-4365/ad4484},
archivePrefix = {arXiv},
       eprint = {2405.14786},
 primaryClass = {astro-ph.EP},
       adsurl = {https://ui.adsabs.harvard.edu/abs/2024ApJS..272...32P}
}

@ARTICLE{2018ARA&A..56..175D,
       author = {{Dawson}, Rebekah I. and {Johnson}, John Asher},
        title = "{Origins of Hot Jupiters}",
      journal = {\araa},
         year = 2018,
        month = sep,
       volume = {56},
        pages = {175-221},
          doi = {10.1146/annurev-astro-081817-051853},
archivePrefix = {arXiv},
       eprint = {1801.06117},
 primaryClass = {astro-ph.EP},
       adsurl = {https://ui.adsabs.harvard.edu/abs/2018ARA&A..56..175D}
}

@ARTICLE{dumusque2011one,
       author = {{Dumusque}, X. and {Udry}, S. and {Lovis}, C. and {Santos}, N.~C. and
         {Monteiro}, M.~J.~P.~F.~G.},
        title = "{Planetary detection limits taking into account stellar noise. I. Observational strategies to reduce stellar oscillation and granulation effects}",
      journal = {\aap},
         year = 2011,
        month = jan,
       volume = {525},
          eid = {A140},
        pages = {A140},
          doi = {10.1051/0004-6361/201014097},
archivePrefix = {arXiv},
       eprint = {1010.2616},
 primaryClass = {astro-ph.EP},
       adsurl = {https://ui.adsabs.harvard.edu/abs/2011A&A...525A.140D}
}

@ARTICLE{1984ApJ...287..769N,
       author = {{Noyes}, R.~W. and {Weiss}, N.~O. and {Vaughan}, A.~H.},
        title = "{The relation between stellar rotation rate and activity cycle periods.}",
      journal = {\apj},
         year = 1984,
        month = dec,
       volume = {287},
        pages = {769-773},
          doi = {10.1086/162735},
       adsurl = {https://ui.adsabs.harvard.edu/abs/1984ApJ...287..769N}
}

@ARTICLE{1963ApJ...138..832W,
       author = {{Wilson}, O.~C.},
        title = "{A Probable Correlation Between Chromospheric Activity and Age in Main-Sequence Stars.}",
      journal = {\apj},
         year = 1963,
        month = oct,
       volume = {138},
        pages = {832},
          doi = {10.1086/147689},
       adsurl = {https://ui.adsabs.harvard.edu/abs/1963ApJ...138..832W}
}

@ARTICLE{2009A&A...499L...9P,
       author = {{Pace}, G. and {Melendez}, J. and {Pasquini}, L. and {Carraro}, G. and {Danziger}, J. and {Fran{\c{c}}ois}, P. and {Matteucci}, F. and {Santos}, N.~C.},
        title = "{An investigation of chromospheric activity spanning the Vaughan-Preston gap: impact on stellar ages}",
      journal = {\aap},
         year = 2009,
        month = may,
       volume = {499},
       number = {1},
        pages = {L9-L12},
          doi = {10.1051/0004-6361/200912090},
archivePrefix = {arXiv},
       eprint = {0904.3199},
 primaryClass = {astro-ph.SR},
       adsurl = {https://ui.adsabs.harvard.edu/abs/2009A&A...499L...9P}
}

@ARTICLE{2008ApJ...687.1264M,
       author = {{Mamajek}, Eric E. and {Hillenbrand}, Lynne A.},
        title = "{Improved Age Estimation for Solar-Type Dwarfs Using Activity-Rotation Diagnostics}",
      journal = {\apj},
         year = 2008,
        month = nov,
       volume = {687},
       number = {2},
        pages = {1264-1293},
          doi = {10.1086/591785},
archivePrefix = {arXiv},
       eprint = {0807.1686},
 primaryClass = {astro-ph},
       adsurl = {https://ui.adsabs.harvard.edu/abs/2008ApJ...687.1264M}
}

@INPROCEEDINGS{dumusque2011two,
       author = {{Dumusque}, Xavier and {Lovis}, Cristophe and {Udry}, Stephane and {Santos}, Nuno C.},
        title = "{Stellar noise and planet detection. II. Radial-velocity noise induced by magnetic cycles}",
    booktitle = {The Astrophysics of Planetary Systems: Formation, Structure, and Dynamical Evolution},
         year = 2011,
       editor = {{Sozzetti}, Alessandro and {Lattanzi}, Mario G. and {Boss}, Alan P.},
       series = {IAU Symposium},
       volume = {276},
        month = nov,
        pages = {530-532},
          doi = {10.1017/S1743921311021090},
archivePrefix = {arXiv},
       eprint = {1011.5581},
 primaryClass = {astro-ph.EP},
       adsurl = {https://ui.adsabs.harvard.edu/abs/2011IAUS..276..530D}
}

@ARTICLE{2012MNRAS.419.2683G,
       author = {{Gibson}, N.~P. and {Aigrain}, S. and {Roberts}, S. and {Evans}, T.~M. and {Osborne}, M. and {Pont}, F.},
        title = "{A Gaussian process framework for modelling instrumental systematics: application to transmission spectroscopy}",
      journal = {\mnras},
         year = 2012,
        month = jan,
       volume = {419},
       number = {3},
        pages = {2683-2694},
          doi = {10.1111/j.1365-2966.2011.19915.x},
archivePrefix = {arXiv},
       eprint = {1109.3251},
 primaryClass = {astro-ph.EP},
       adsurl = {https://ui.adsabs.harvard.edu/abs/2012MNRAS.419.2683G}
}

@ARTICLE{2014ApJ...789..154D,
       author = {{Dumusque}, Xavier and {Bonomo}, Aldo S. and {Haywood}, Rapha{\"e}lle D. and {Malavolta}, Luca and {S{\'e}gransan}, Damien and {Buchhave}, Lars A. and {Collier Cameron}, Andrew and {Latham}, David W. and {Molinari}, Emilio and {Pepe}, Francesco and {Udry}, St{\'e}phane and {Charbonneau}, David and {Cosentino}, Rosario and {Dressing}, Courtney D. and {Figueira}, Pedro and {Fiorenzano}, Aldo F.~M. and {Gettel}, Sara and {Harutyunyan}, Avet and {Horne}, Keith and {Lopez-Morales}, Mercedes and {Lovis}, Christophe and {Mayor}, Michel and {Micela}, Giusi and {Motalebi}, Fatemeh and {Nascimbeni}, Valerio and {Phillips}, David F. and {Piotto}, Giampaolo and {Pollacco}, Don and {Queloz}, Didier and {Rice}, Ken and {Sasselov}, Dimitar and {Sozzetti}, Alessandro and {Szentgyorgyi}, Andrew and {Watson}, Chris},
        title = "{The Kepler-10 Planetary System Revisited by HARPS-N: A Hot Rocky World and a Solid Neptune-Mass Planet}",
      journal = {\apj},
         year = 2014,
        month = jul,
       volume = {789},
       number = {2},
          eid = {154},
        pages = {154},
          doi = {10.1088/0004-637X/789/2/154},
archivePrefix = {arXiv},
       eprint = {1405.7881},
 primaryClass = {astro-ph.EP},
       adsurl = {https://ui.adsabs.harvard.edu/abs/2014ApJ...789..154D}
}

@ARTICLE{2025MNRAS.539.2230R,
       author = {{Rogers}, James G.},
        title = "{On the road to the radius valley: distinguishing between gas dwarfs and water worlds with young transiting exoplanets}",
      journal = {\mnras},
         year = 2025,
        month = may,
       volume = {539},
       number = {3},
        pages = {2230-2241},
          doi = {10.1093/mnras/staf628},
archivePrefix = {arXiv},
       eprint = {2503.17364},
 primaryClass = {astro-ph.EP},
       adsurl = {https://ui.adsabs.harvard.edu/abs/2025MNRAS.539.2230R}
}

@ARTICLE{2015ApJ...810..136G,
       author = {{Genda}, H. and {Kobayashi}, H. and {Kokubo}, E.},
        title = "{Warm Debris Disks Produced by Giant Impacts during Terrestrial Planet Formation}",
      journal = {\apj},
         year = 2015,
        month = sep,
       volume = {810},
       number = {2},
          eid = {136},
        pages = {136},
          doi = {10.1088/0004-637X/810/2/136},
archivePrefix = {arXiv},
       eprint = {1508.00977},
 primaryClass = {astro-ph.EP},
       adsurl = {https://ui.adsabs.harvard.edu/abs/2015ApJ...810..136G}
}

@ARTICLE{2025ApJ...989..117K,
       author = {{Kimura}, Tadahiro and {Kokubo}, Eiichiro and {Matsumoto}, Yuji and {Mordasini}, Christoph and {Ikoma}, Masahiro},
        title = "{Semianalytical Model for the Dynamical Evolution of Planetary Systems. II. Application to Systems Formed by a Planet Formation Model}",
      journal = {\apj},
         year = 2025,
        month = aug,
       volume = {989},
       number = {1},
          eid = {117},
        pages = {117},
          doi = {10.3847/1538-4357/adef15},
archivePrefix = {arXiv},
       eprint = {2507.03906},
 primaryClass = {astro-ph.EP},
       adsurl = {https://ui.adsabs.harvard.edu/abs/2025ApJ...989..117K}
}

@ARTICLE{2019A&A...630A..81B,
       author = {{Benatti}, S. and {Nardiello}, D. and {Malavolta}, L. and {Desidera}, S. and {Borsato}, L. and {Nascimbeni}, V. and {Damasso}, M. and {D'Orazi}, V. and {Mesa}, D. and {Messina}, S. and {Esposito}, M. and {Bignamini}, A. and {Claudi}, R. and {Covino}, E. and {Lovis}, C. and {Sabotta}, S.},
        title = "{A possibly inflated planet around the bright young star DS Tucanae A}",
      journal = {\aap},
         year = 2019,
        month = oct,
       volume = {630},
          eid = {A81},
        pages = {A81},
          doi = {10.1051/0004-6361/201935598},
archivePrefix = {arXiv},
       eprint = {1904.01591},
 primaryClass = {astro-ph.SR},
       adsurl = {https://ui.adsabs.harvard.edu/abs/2019A&A...630A..81B}
}

@ARTICLE{2016AJ....152..204L,
       author = {{L{\'o}pez-Morales}, Mercedes and {Haywood}, Rapha{\"e}lle D. and {Coughlin}, Jeffrey L. and {Zeng}, Li and {Buchhave}, Lars A. and {Giles}, Helen A.~C. and {Affer}, Laura and {Bonomo}, Aldo S. and {Charbonneau}, David and {Collier Cameron}, Andrew and {Consentino}, Rosario and {Dressing}, Courtney D. and {Dumusque}, Xavier and {Figueira}, Pedro and {Fiorenzano}, Aldo F.~M. and {Harutyunyan}, Avet and {Johnson}, John Asher and {Latham}, David W. and {Lopez}, Eric D. and {Lovis}, Christophe and {Malavolta}, Luca and {Mayor}, Michel and {Micela}, Giusi and {Molinari}, Emilio and {Mortier}, Annelies and {Motalebi}, Fatemeh and {Nascimbeni}, Valerio and {Pepe}, Francesco and {Phillips}, David F. and {Piotto}, Giampaolo and {Pollacco}, Don and {Queloz}, Didier and {Rice}, Ken and {Sasselov}, Dimitar and {Segransan}, Damien and {Sozzetti}, Alessandro and {Udry}, Stephane and {Vanderburg}, Andrew and {Watson}, Chris},
        title = "{Kepler-21b: A Rocky Planet Around a V = 8.25 Magnitude Star}",
      journal = {\aj},
         year = 2016,
        month = dec,
       volume = {152},
       number = {6},
          eid = {204},
        pages = {204},
          doi = {10.3847/0004-6256/152/6/204},
archivePrefix = {arXiv},
       eprint = {1609.07617},
 primaryClass = {astro-ph.EP},
       adsurl = {https://ui.adsabs.harvard.edu/abs/2016AJ....152..204L}
}

@ARTICLE{2017A&A...608A..25B,
       author = {{Barros}, S.~C.~C. and {Gosselin}, H. and {Lillo-Box}, J. and {Bayliss}, D. and {Delgado Mena}, E. and {Brugger}, B. and {Santerne}, A. and {Armstrong}, D.~J. and {Adibekyan}, V. and {Armstrong}, J.~D. and {Barrado}, D. and {Bento}, J. and {Boisse}, I. and {Bonomo}, A.~S. and {Bouchy}, F. and {Brown}, D.~J.~A. and {Cochran}, W.~D. and {Collier Cameron}, A. and {Deleuil}, M. and {Demangeon}, O. and {D{\'\i}az}, R.~F. and {Doyle}, A. and {Dumusque}, X. and {Ehrenreich}, D. and {Espinoza}, N. and {Faedi}, F. and {Faria}, J.~P. and {Figueira}, P. and {Foxell}, E. and {H{\'e}brard}, G. and {Hojjatpanah}, S. and {Jackman}, J. and {Lendl}, M. and {Ligi}, R. and {Lovis}, C. and {Melo}, C. and {Mousis}, O. and {Neal}, J.~J. and {Osborn}, H.~P. and {Pollacco}, D. and {Santos}, N.~C. and {Sefako}, R. and {Shporer}, A. and {Sousa}, S.~G. and {Triaud}, A.~H.~M.~J. and {Udry}, S. and {Vigan}, A. and {Wyttenbach}, A.},
        title = "{Precise masses for the transiting planetary system HD 106315 with HARPS}",
      journal = {\aap},
         year = 2017,
        month = dec,
       volume = {608},
          eid = {A25},
        pages = {A25},
          doi = {10.1051/0004-6361/201731276},
archivePrefix = {arXiv},
       eprint = {1709.00865},
 primaryClass = {astro-ph.EP},
       adsurl = {https://ui.adsabs.harvard.edu/abs/2017A&A...608A..25B}
}

@ARTICLE{2020AJ....159...23N,
       author = {{Nava}, Chantanelle and {L{\'o}pez-Morales}, Mercedes and {Haywood}, Rapha{\"e}lle D. and {Giles}, Helen A.~C.},
        title = "{Exoplanet Imitators: A Test of Stellar Activity Behavior in Radial Velocity Signals}",
      journal = {\aj},
         year = 2020,
        month = jan,
       volume = {159},
       number = {1},
          eid = {23},
        pages = {23},
          doi = {10.3847/1538-3881/ab53ec},
archivePrefix = {arXiv},
       eprint = {1911.04106},
 primaryClass = {astro-ph.EP},
       adsurl = {https://ui.adsabs.harvard.edu/abs/2020AJ....159...23N}
}

@ARTICLE{2026arXiv260324682K,
       author = {{Kroft}, Maxwell A. and {Beatty}, Thomas G. and {Salzer}, Joseph M. and {Zwicker}, Claire and {Triantafillides}, Anastasia and {Becker}, Juliette and {Soares-Furtado}, Melinda and {Cisewski-Kehe}, Jessi and {Lissauer}, Jack J. and {Armitage}, Tayt S. and {Livesey}, Joseph R. and {Narayan}, Ritvik Sai and {Widicus Weaver}, Susanna and {Zhang}, Ke and {Bieryla}, Allyson and {Ciardi}, David R. and {Clark}, Catherine A. and {Felsmann}, Miranda and {Fernandes}, Rachel B. and {Howell}, Steve B. and {Lund}, Michael B.},
        title = "{GJ 523b is a Massive, 170 Myr-old Mega-Earth, Likely on a Polar Orbit}",
      journal = {arXiv e-prints},
         year = 2026,
        month = mar,
          eid = {arXiv:2603.24682},
        pages = {arXiv:2603.24682},
          doi = {10.48550/arXiv.2603.24682},
archivePrefix = {arXiv},
       eprint = {2603.24682},
 primaryClass = {astro-ph.EP},
       adsurl = {https://ui.adsabs.harvard.edu/abs/2026arXiv260324682K}
}

@ARTICLE{2019ApJ...875L...7D,
       author = {{Dragomir}, Diana and {Teske}, Johanna and {G{\"u}nther}, Maximilian N. and {S{\'e}gransan}, Damien and {Burt}, Jennifer A. and {Huang}, Chelsea X. and {Vanderburg}, Andrew and {Matthews}, Elisabeth and {Dumusque}, Xavier and {Stassun}, Keivan G. and {Pepper}, Joshua and {Ricker}, George R. and {Vanderspek}, Roland and {Latham}, David W. and {Seager}, Sara and {Winn}, Joshua N. and {Jenkins}, Jon M. and {Beatty}, Thomas and {Bouchy}, Fran{\c{c}}ois and {Brown}, Timothy M. and {Butler}, R. Paul and {Ciardi}, David R. and {Crane}, Jeffrey D. and {Eastman}, Jason D. and {Fossati}, Luca and {Francis}, Jim and {Fulton}, Benjamin J. and {Gaudi}, B. Scott and {Goeke}, Robert F. and {James}, David and {Klaus}, Todd C. and {Kuhn}, Rudolf B. and {Lovis}, Christophe and {Lund}, Michael B. and {McDermott}, Scott and {Paegert}, Martin and {Pepe}, Francesco and {Rodriguez}, Joseph E. and {Sha}, Lizhou and {Shectman}, Stephen A. and {Shporer}, Avi and {Siverd}, Robert J. and {Garcia Soto}, Aylin and {Stevens}, Daniel J. and {Twicken}, Joseph D. and {Udry}, St{\'e}phane and {Villanueva}, Jr., Steven and {Wang}, Sharon X. and {Wohler}, Bill and {Yao}, Xinyu and {Zhan}, Zhuchang},
        title = "{TESS Delivers Its First Earth-sized Planet and a Warm Sub-Neptune}",
      journal = {\apjl},
         year = 2019,
        month = apr,
       volume = {875},
       number = {2},
          eid = {L7},
        pages = {L7},
          doi = {10.3847/2041-8213/ab12ed},
archivePrefix = {arXiv},
       eprint = {1901.00051},
 primaryClass = {astro-ph.EP},
       adsurl = {https://ui.adsabs.harvard.edu/abs/2019ApJ...875L...7D}
}

@ARTICLE{boro2018,
       author = {{Boro Saikia}, S. and {Marvin}, C.~J. and {Jeffers}, S.~V. and {Reiners}, A. and {Cameron}, R. and {Marsden}, S.~C. and {Petit}, P. and {Warnecke}, J. and {Yadav}, A.~P.},
        title = "{Chromospheric activity catalogue of 4454 cool stars. Questioning the active branch of stellar activity cycles}",
      journal = {\aap},
         year = 2018,
        month = aug,
       volume = {616},
          eid = {A108},
        pages = {A108},
          doi = {10.1051/0004-6361/201629518},
archivePrefix = {arXiv},
       eprint = {1803.11123},
 primaryClass = {astro-ph.SR},
       adsurl = {https://ui.adsabs.harvard.edu/abs/2018A&A...616A.108B}
}

@ARTICLE{rosario24,
       author = {{Ros{\'a}rio}, N.~M. and {Demangeon}, O.~D.~S. and {Barros}, S.~C.~C. and {Gandolfi}, D. and {Egger}, J.~A. and {Serrano}, L.~M. and {Osborn}, H.~P. and {Beck}, M. and {Benz}, W. and {Flor{\'e}n}, H.-G. and {Guterman}, P. and {Wilson}, T.~G. and {Alibert}, Y. and {Fossati}, L. and {Hooton}, M.~J. and {Delrez}, L. and {Santos}, N.~C. and {Sousa}, S.~G. and {Bonfanti}, A. and {Salmon}, S. and {Adibekyan}, V. and {Nigioni}, A. and {Venturini}, J. and {Alonso}, R. and {Anglada}, G. and {Asquier}, J. and {B{\'a}rczy}, T. and {Barrado Navascues}, D. and {Barrag{\'a}n}, O. and {Baumjohann}, W. and {Beck}, T. and {Billot}, N. and {Biondi}, F. and {Bonfils}, X. and {Borsato}, L. and {Brandeker}, A. and {Broeg}, C. and {Cessa}, V. and {Charnoz}, S. and {Collier Cameron}, A. and {Csizmadia}, Sz. and {Cubillos}, P.~E. and {Davies}, M.~B. and {Deleuil}, M. and {Deline}, A. and {Demory}, B.-O. and {Ehrenreich}, D. and {Erikson}, A. and {Esposito}, M. and {Fortier}, A. and {Fridlund}, M. and {Gillon}, M. and {G{\"u}del}, M. and {G{\"u}nther}, M.~N. and {Helling}, Ch. and {Hoyer}, S. and {Isaak}, K.~G. and {Kiss}, L.~L. and {Lam}, K.~W.~F. and {Laskar}, J. and {Lecavelier des Etangs}, A. and {Lendl}, M. and {Luntzer}, A. and {Magrin}, D. and {Maxted}, P.~F.~L. and {Mordasini}, C. and {Nascimbeni}, V. and {Olofsson}, G. and {Osborne}, H.~L.~M. and {Ottensamer}, R. and {Pagano}, I. and {Pall{\'e}}, E. and {Peter}, G. and {Piotto}, G. and {Pollacco}, D. and {Queloz}, D. and {Ragazzoni}, R. and {Rando}, N. and {Rauer}, H. and {Ribas}, I. and {Scandariato}, G. and {S{\'e}gransan}, D. and {Simon}, A.~E. and {Smith}, A.~M.~S. and {Stalport}, M. and {Szab{\'o}}, Gy. M. and {Thomas}, N. and {Udry}, S. and {Van Eylen}, V. and {Van Grootel}, V. and {Villaver}, E. and {Walter}, I. and {Walton}, N.~A.},
        title = "{Precise characterisation of HD 15337 with CHEOPS: A laboratory for planet formation and evolution}",
      journal = {\aap},
         year = 2024,
        month = jun,
       volume = {686},
          eid = {A282},
        pages = {A282},
          doi = {10.1051/0004-6361/202347759},
archivePrefix = {arXiv},
       eprint = {2403.16621},
 primaryClass = {astro-ph.EP},
       adsurl = {https://ui.adsabs.harvard.edu/abs/2024A&A...686A.282R}
}

@ARTICLE{2024ApJ...975L..22W,
       author = {{Weiner Mansfield}, Megan and {Xue}, Qiao and {Zhang}, Michael and {Mahajan}, Alexandra S. and {Ih}, Jegug and {Koll}, Daniel and {Bean}, Jacob L. and {Coy}, Brandon Park and {Eastman}, Jason D. and {Kempton}, Eliza M.-R. and {Kite}, Edwin S.},
        title = "{No Thick Atmosphere on the Terrestrial Exoplanet Gl 486b}",
      journal = {\apjl},
         year = 2024,
        month = nov,
       volume = {975},
       number = {1},
          eid = {L22},
        pages = {L22},
          doi = {10.3847/2041-8213/ad8161},
archivePrefix = {arXiv},
       eprint = {2408.15123},
 primaryClass = {astro-ph.EP},
       adsurl = {https://ui.adsabs.harvard.edu/abs/2024ApJ...975L..22W}
}

@ARTICLE{2026arXiv260618355L,
       author = {{Lee}, Rena A. and {Dai}, Fei and {Price}, Ellen M. and {Han}, Te and {Gandolfi}, Davide and {Zechmeister}, Mathias and {Stef{\'a}nsson}, Gu{\dh}mundur and {Dong}, Jiayin and {Albrecht}, Simon H. and {Lam}, Kristine W.~F. and {Chiti}, Federica and {van Saders}, Jennifer L. and {Huber}, Daniel and {Knutson}, Heather A. and {Collins}, Karen A. and {Zhang}, Michael and {Rogers}, Leslie A. and {Armano}, Eleonora and {Brinkman}, Casey L. and {Saunders}, Nicholas and {Hey}, Daniel},
        title = "{Revisiting the Exo-Mercury Candidate GJ 367 b with ESPRESSO and a Self-Consistent Tidal Distortion Model}",
      journal = {arXiv e-prints},
         year = 2026,
        month = jun,
          eid = {arXiv:2606.18355},
        pages = {arXiv:2606.18355},
          doi = {10.48550/arXiv.2606.18355},
archivePrefix = {arXiv},
       eprint = {2606.18355},
 primaryClass = {astro-ph.EP},
       adsurl = {https://ui.adsabs.harvard.edu/abs/2026arXiv260618355L}
}

@ARTICLE{weisserman26,
       author = {{Weisserman}, Drew and {Gromek}, Nicole and {Cloutier}, Ryan and {Bali}, Komal and {Cadieux}, Charles and {Plotnykov}, Mykhaylo and {L'Heureux}, Alexandrine and {Srivastava}, Avidaan and {Carmona}, Andres and {Frensch}, Yolanda G.~C. and {Artigau}, {\'E}tienne and {Baron}, Fr{\'e}d{\'e}rique and {Barros}, Susana C.~C. and {Benneke}, Bj{\"o}rn and {Bonfils}, Xavier and {Bouchy}, Fran{\c{c}}ois and {Bryan}, Marta and {Cook}, Neil J. and {Cowan}, Nicolas B. and {Cristo}, Eduardo and {Delfosse}, Xavier and {Doyon}, Ren{\'e} and {Dumusque}, Xavier and {Ehrenreich}, David and {Gonz{\'a}lez Hern{\'a}ndez}, Jonay I. and {Lafreni{\`e}re}, David and {de Castro Le{\~a}o}, Izan and {Lovis}, Christophe and {Malo}, Lison and {Canto Martins}, Bruno L. and {Su{\'a}rez Mascare{\~n}o}, Alejandro and {De Medeiros}, Jose Renan and {Melo}, Claudio and {Mignon}, Lucile and {Mordasini}, Christoph and {Pepe}, Francesco and {Rebolo}, Rafael and {Rowe}, Jason and {Santos}, Nuno C. and {S{\'e}gransan}, Damien and {Udry}, St{\'e}phane and {Valencia}, Diana and {Wade}, Gregg and {Aguiar}, Jos{\'e} Luan A. and {Allart}, Romain and {Bazinet}, Luc and {Delisle}, Jean-Baptiste and {B{\'e}langer}, Flavie and {Blackman}, Joshua and {Bourrier}, Vincent and {Branco}, Pedro and {Bruniquel}, Vincent and {Carteret}, Yann and {Cointepas}, Marion and {Darveau-Bernier}, Antoine and {Dauplaise}, Laurie and {Delgado-Mena}, Elisa and {Dorn}, Caroline and {Doshi}, Dhvani and {Faria}, Jo{\~a}o and {Fontinele}, Dasaev O. and {Forveille}, Thierry and {Gagn{\'e}}, Jonathan and {Genest}, Fr{\'e}d{\'e}ric and {Glover}, Jennifer and {de Lima Gomes}, Roseane and {Grieves}, Nolan and {Hobson}, Melissa J. and {Hoeijmakers}, H. Jens and {Jahandar}, Farbod and {Krishnamurthy}, Vigneshwaran and {Lamontagne}, Pierrot and {Larue}, Pierre and {Leath}, Henry and {Lim}, Olivia and {Lipper}, Justin and {Messamah}, Lina and {Messias}, Yuri S. and {Monteiro}, Telmo and {Moranta}, Leslie and {Al Moulla}, Khaled and {Mounzer}, Dany and {Mraz}, Georgia and {Nari}, Nicola and {Nielsen}, Louise D. and {Osborn}, Ares and {Otegi}, Jon and {Parc}, L{\'e}na and {Pelletier}, Stefan and {Pereira}, Olivia and {Piaulet-Ghorayeb}, Caroline and {Rosener}, Riley and {Seidel}, Julia and {Gomes da Silva}, Jo{\~a}o and {Costa Silva}, Ana Rita and {Stefanov}, Atanas K. and {Teixeira}, M{\'a}rcio A. and {Vandal}, Thomas and {Vaulato}, Valentina and {Wardenier}, Joost P. and {Yariv}, Vincent},
        title = "{Super-Earth masses and stellar abundances from NIRPS reveal tentative evidence for water-rich formation around M dwarfs}",
      journal = {\aap},
         year = 2026,
        month = may,
       volume = {709},
          eid = {A165},
        pages = {A165},
          doi = {10.1051/0004-6361/202558011},
archivePrefix = {arXiv},
       eprint = {2604.07447},
 primaryClass = {astro-ph.EP},
       adsurl = {https://ui.adsabs.harvard.edu/abs/2026A&A...709A.165W}
}

@ARTICLE{2024ApJ...973L...8X,
       author = {{Xue}, Qiao and {Bean}, Jacob L. and {Zhang}, Michael and {Mahajan}, Alexandra and {Ih}, Jegug and {Eastman}, Jason D. and {Lunine}, Jonathan and {Mansfield}, Megan Weiner and {Coy}, Brandon Park and {Kempton}, Eliza M.-R. and {Koll}, Daniel and {Kite}, Edwin},
        title = "{JWST Thermal Emission of the Terrestrial Exoplanet GJ 1132b}",
      journal = {\apjl},
         year = 2024,
        month = sep,
       volume = {973},
       number = {1},
          eid = {L8},
        pages = {L8},
          doi = {10.3847/2041-8213/ad72e9},
archivePrefix = {arXiv},
       eprint = {2408.13340},
 primaryClass = {astro-ph.EP},
       adsurl = {https://ui.adsabs.harvard.edu/abs/2024ApJ...973L...8X}
}

@ARTICLE{1996AJ....111..439H,
       author = {{Henry}, Todd J. and {Soderblom}, David R. and {Donahue}, Robert A. and {Baliunas}, Sallie L.},
        title = "{A Survey of Ca II H and K Chromospheric Emission in Southern Solar-Type Stars}",
      journal = {\aj},
         year = 1996,
        month = jan,
       volume = {111},
        pages = {439},
          doi = {10.1086/117796},
       adsurl = {https://ui.adsabs.harvard.edu/abs/1996AJ....111..439H}
}

@ARTICLE{2024NatSR..1427219L,
       author = {{Livingston}, John H. and {Gandolfi}, Davide and {Trani}, Alessandro A. and {Herath}, Mahesh and {Barrag{\'a}n}, Oscar and {Hatzes}, Artie and {Luque}, Rafael and {Fukui}, Akihiko and {Nowak}, Grzegorz and {Palle}, Enric and {Hellier}, Coel and {Fridlund}, Malcolm and {de Leon}, Jerome and {Hirano}, Teruyuki and {Narita}, Norio and {Albrecht}, Simon and {Dai}, Fei and {Deeg}, Hans and {Van Eylen}, Vincent and {Korth}, Judith and {Tamura}, Motohide},
        title = "{An ultra-short-period super-Earth with an extremely high density and an outer companion}",
      journal = {Scientific Reports},
         year = 2024,
        month = nov,
       volume = {14},
       number = {1},
          eid = {27219},
        pages = {27219},
          doi = {10.1038/s41598-024-76490-y},
       adsurl = {https://ui.adsabs.harvard.edu/abs/2024NatSR..1427219L}
}

@ARTICLE{2024MNRAS.529..141G,
       author = {{Guenther}, E.~W. and {Goffo}, E. and {Sebastian}, D. and {Smith}, A.~M.~S. and {Persson}, C.~M. and {Fridlund}, M. and {Gandolfi}, D. and {Korth}, J.},
        title = "{A new mass and radius determination of the ultra-short period planet K2-106b and the fluffy planet K2-106c}",
      journal = {\mnras},
         year = 2024,
        month = mar,
       volume = {529},
       number = {1},
        pages = {141-154},
          doi = {10.1093/mnras/stae494},
archivePrefix = {arXiv},
       eprint = {2402.09322},
 primaryClass = {astro-ph.EP},
       adsurl = {https://ui.adsabs.harvard.edu/abs/2024MNRAS.529..141G}
}
\bibliographystyle{aasjournal}



\end{document}